\documentclass[%
 reprint,
 superscriptaddress,
 aps,
 pra,
 longbibliography,
 floatfix
]{revtex4-2}

\usepackage[english]{babel}
\usepackage{graphicx}
\usepackage{dcolumn}
\usepackage{bm}

\usepackage{multirow}
\usepackage{lineno}
\usepackage{mathrsfs}
\usepackage{amsmath}
\usepackage{braket}
\usepackage{enumitem}
\usepackage{amsfonts}
\usepackage{xcolor}
\usepackage{siunitx}
\usepackage{booktabs}
\usepackage{microtype}
\usepackage{xurl}
\usepackage[colorlinks=true,allcolors=blue]{hyperref}

\microtypesetup{nopatch=footnote}
\begin{document}

\preprint{APS/123-QED}

\title{Simulation-guided design of an integrated photonic cavity for frequency-multiplexed spontaneous parametric down-conversion}

\author{Benjamin Szamosfalvi}
 \email{Contact author: andeloth@byu.edu}
 \affiliation{Department of Electrical and Computer Engineering, Brigham Young University, Provo, Utah, USA}
 \author{Michael Raymer}
 \affiliation{Department of Physics and Oregon Center for Optical, Molecular, and Quantum Science, University of Oregon, Eugene, Oregon, USA}
\author{CJ Xin}
 \affiliation{Department of Electrical Engineering, Harvard University, Boston, Massachusetts, USA}
\author{Leticia Magalhaes}
 \affiliation{Department of Electrical Engineering, Harvard University, Boston, Massachusetts, USA}
\author{Jarrett Nelson}
 \affiliation{Department of Electrical and Computer Engineering, Brigham Young University, Provo, Utah, USA}
\author{Marko Lončar}
 \affiliation{Department of Electrical Engineering, Harvard University, Boston, Massachusetts, USA}
\author{Ryan M. Camacho}
 \email{Contact author: camacho@byu.edu}
 \affiliation{Department of Electrical and Computer Engineering, Brigham Young University, Provo, Utah, USA}

\date{\today}

\begin{abstract}
Frequency-multiplexed entangled photon pair sources with narrow bandwidths and high pair generation efficiency are a key enabling technology for quantum networking. We present a simulation-based design study of an integrated photonic racetrack resonator source for spontaneous parametric down-conversion (SPDC) that simultaneously achieves all three properties. The central result is a simulated set of 106 doubly resonant signal and idler frequency-mode pairs with an effective frequency-space Schmidt number of 105.99, an average single-island purity of 0.994, an average bandwidth of \SI{0.513}{\giga\hertz}, a mean free spectral range of \SI{46.9}{\giga\hertz}, and a total internal pair-generation-rate efficiency of \SI{1.75}{\giga\hertz\per\milli\watt}. Under deterministic wavelength-based splitting, the folded 53-pair state has an effective frequency-space Schmidt number of 53.00. To support these predictions, we derive a closed-form analytical connection between classical cavity parameters (resonant frequencies, decay rates, and coupling coefficients) and the quantum joint spectral amplitude and pair generation rate, extending the dispersive-medium quantization formalism of Raymer to the nonlinear optical cavity case. We demonstrate how classical electromagnetic field simulations can be combined with this analytical framework to predict quantum figures of merit for an integrated photonic source prior to fabrication. Fabrication and experimental validation are left for future work.
\end{abstract}

\maketitle

\section{Introduction}

\begin{table*}[th]
    \centering
    \caption{Comparison of performance metrics of integrated cavity-based entangled photon pair sources. For the case of multiplexed sources, the pair generation rate listed is the sum of individual resonances' PGRs and the bandwidth is the average bandwidth across resonances. All PGR values refer to the rate of photon pairs produced inside the resonator source before coupling losses, waveguide and fiber propagation losses, or detector losses are considered.}\label{tab:sourceperf-comparison}
    \begin{ruledtabular}
        \begin{tabular}{lcdccll}
            Reference &\multicolumn{1}{c}{Internal PGR/mW} & \multicolumn{1}{c}{Brightness\footnotemark[1]} & \multicolumn{1}{c}{Bandwidth} &  Schmidt number in $\omega$ & Platform & Date\\
        \midrule
            Guo \textit{et al.}~\cite{Guo2017} & \SI{5.80}{\mega\hertz} & 5.27 & \SI{1.10}{\giga\hertz} & 1 & AlN & 2016\\
            Imany \textit{et al.}~\cite{Imany-2018-sfwm-highschmidtN} & N/A & \multicolumn{1}{c}{\text{N/A}} & \SI{100}{\mega\hertz} & 20 & SiN & 2018\\
            Steiner \textit{et al.}~\cite{steiner-narrowband-cavity-spdc-bright} & \SI{20.0}{\giga\hertz} & 1.29 \times 10^5 & \SI{155}{\mega\hertz} & 1 & AlGaAs & 2021\\
            Mahmudlu \textit{et al.}~\cite{onchip-lased-multiplexed-swfm-mahmudlu} &  \SI{8.41}{\kilo\hertz} & 1.29 \times 10^{-3} & \SI{1.63}{\giga\hertz} & 4\footnotemark[2] & SiN & 2023\\
            Zhang \textit{et al.}~\cite{zhang-multiplexed-nonpoled-spdc} &  \SI{20.0}{\mega\hertz} & 3.23 \times 10^1 & \SI{0.620}{\giga\hertz} & 9.2 & TFLN  & 2023\\
            Henry \textit{et al.}~\cite{Henry-2023-multiplexed-source} & \SI{21.2}{\mega\hertz} & 5.05 \times 10^{-1} & \SI{0.600}{\giga\hertz} & 70\footnotemark[2] & Si & 2023 \\
            Chen \textit{et al.}~\cite{chen-narrowband-sfwm-integrated-cavity} & \SI{30.3}{\mega\hertz} & 1.17 \times 10^3 & \SI{25.9}{\mega\hertz} & 1 & SiN & 2024 \\
            Pang \textit{et al.}~\cite{sfwm-multiplexed-algaas-pang} &  \SI{280}{\mega\hertz} (\SI{5.1}{\giga\hertz})\footnotemark[3] & 2.46 \times 10^3\footnotemark[3] & \SI{0.630}{\giga\hertz}\footnotemark[3] & 5\footnotemark[2] (1) & AlGaAs & 2025\\
            This work & \SI{1.75}{\giga\hertz} (\SI{1.75}{\giga\hertz})\footnotemark[4] & 1.85 \times 10^1 & \SI{0.513}{\giga\hertz} & 105.99 & TFLN\footnotemark[4] & 2026\\
        \end{tabular}
    \end{ruledtabular}
    \footnotetext[1]{Brightness is listed in units of $10^6$ pairs per second per milliwatt of pump power per GHz of frequency bandwidth (MHz/mW/GHz).}
    \footnotetext[2]{Schmidt number not explicitly stated, number of joint frequency modes listed instead.}
    \footnotetext[3]{Multiple parallelized resonators each produce photon pairs at \SI{1}{\giga\hertz\per\milli\watt} internal efficiency, but when parallelized can produce frequency-entangled photon pairs at \SI{280}{\mega\hertz\per\milli\watt}. For Pang's source, the brightness value is calculated for each listed resonance in Pang \textit{et al.} Table I, then averaged since the bandwidths vary significantly between resonances.} 
    \footnotetext[4]{Specific performance predictions for this work are for an example device on a TFLN platform, but our simulations can easily be adapted for different materials. The \SI{1.75}{\giga\hertz\per\milli\watt} total PGR is the sum across all 106 islands; the average per-island PGR is approximately \SI{16.55}{\mega\hertz\per\milli\watt} with a standard deviation of \SI{0.042}{\mega\hertz}, a maximum single-island PGR of \SI{16.63}{\mega\hertz\per\milli\watt}, and a minimum single-island PGR of \SI{16.49}{\mega\hertz\per\milli\watt}. This means each mode is within the usable PGR regime. If artificial losses are included to even out per-mode PGR, an effective PGR of \SI{1.75}{\giga\hertz\per\milli\watt} can be achieved.}
\end{table*}

Distributing quantum entanglement between distant locations is a key goal of quantum technology research. The starting point of entanglement distribution protocols is an entangled photon pair source which must fulfill strict specification targets in order to create distributed entanglement at practically useful rates~\cite{ZALM}. One of the main methods used in classical communications is frequency-multiplexing, which increases data transmission speeds, scaling linearly with bandwidth. A practically useful frequency-multiplexed entangled photon pair source is highly desirable as the starting point of frequency-multiplexed quantum networking schemes. An integrated photonic highly frequency-multiplexed entangled photon pair source would be best suited for large-scale deployment in future quantum network routing ``stations,'' but testing and troubleshooting integrated devices is difficult compared with free-space optics, so extensive simulations need to be run to inform the design process when complex specifications must be met. There are several mature simulation software tools available for performing classical field simulations on photonic devices, but these solvers do not have the capability to solve for certain quantum properties of light such as the joint spectral amplitude (JSA), which describes the joint frequency state of the produced photon pairs. As a consequence, photonic designers cannot fully predict how their source will perform in a quantum networking system by using only classical simulation tools. In this paper, we address this gap with two contributions. First, we derive a closed-form analytical connection between the classical resonator parameters of a cavity SPDC source---resonant frequencies, total decay rates, and coupler transmission coefficients---and the quantum figures of merit: the JSA and pair generation rate (PGR). This derivation uses the dispersive-medium field quantization formalism of Raymer~\cite{Raymer20-dispersive-quantization} as its Hamiltonian starting point---a more appropriate foundation for waveguide sources than the plane-wave treatments used in prior cavity JSA derivations~\cite{jsa-shape-theory-Jeronimo-Moreno2010}---and extends it to the nonlinear optical regime to derive the cavity JSA and PGR. The linear  Hamiltonian is taken from Raymer's framework; a first-principles derivation of the Hamiltonian starting point is deferred to a follow-up paper, but all formulas needed to implement the simulation pipeline are self-contained here. All JSA and PGR predictions use a first-order perturbative SPDC model; at the operating point used here the estimated per-island mean pair number per pump pulse is approximately $1.84 \times 10^{-2}$, which results in few double-pair generation events. Second, we demonstrate how classical full-wave electromagnetic simulations can be combined with this analytical framework to predict the quantum performance of an integrated photonic cavity SPDC source prior to fabrication. We apply this approach to optimize a racetrack resonator design for frequency-entangled, frequency-multiplexed photon pair generation across 106 doubly resonant signal and idler frequency-mode pairs, with an effective frequency-space Schmidt number of 105.99. These figures place the design, to our knowledge, at the high end of reported or predicted integrated on-chip resonator sources effective frequency-state dimensionality which is commonly characterized by the Schmidt number. We note that sources capable of producing several hundreds to over a thousand frequency modes have been demonstrated, but these sources usually show large overlaps between frequency modes which decrease their frequency-space Schmidt numbers and thus limit the number of frequency states that can be heralded from photon pairs they produce~\cite{Yamazaki-2022-manymode-freespace-spdc, Chang-2021-manymode-freespace-spdc, fabre-2020-openfacetwaveguide-multiplexed-source}. Because the literature does not report a single common comparison metric---some papers report raw mode count, others effective mode count, and others Schmidt number---we avoid a sharper ranking claim here and instead use Table~\ref{tab:sourceperf-comparison} to make the comparison basis explicit. In Table~\ref{tab:sourceperf-comparison}, we list only other integrated photonic cavity sources with high quality factors. We note that we do not include any free-space sources in the comparison as free-space source design is very different on the component and experimental level from integrated source design and is therefore outside the scope of this work. Therefore, we make no claims about outperforming state-of-the-art free-space sources in terms of PGR efficiency or brightness, though to the best of the authors' knowledge no free-space source has been demonstrated with more than 106 frequency-space Schmidt modes at the time of writing this work. All results in this paper are simulation predictions; fabrication and experimental validation are left for future work.

Frequency-multiplexed photon pairs are highly desirable for a quantum network due to the straightforward increase in qubit transmission capacity in exchange for requiring frequency demultiplexing before detection~\cite{ZALM, multiplexed-networking-ruskuc2025, multiplexed-networking-fan2025}. The simplest way to frequency-multiplex entangled photon pairs is by combining the output of several sources tuned to different frequency modes~\cite{sfwm-multiplexed-algaas-pang}. However, the use of multiple sources increases hardware costs and pumping several sources requires more power than pumping a single source that is optimized for generation of frequency-multiplexed photon pairs. Therefore, it is desirable to use a single source to populate several frequency modes at the same time. Most state-of-the-art sources are only capable of populating fewer than ten frequency mode signal and idler pairs, which would require a significant number of sources to be used in parallel to populate the desired number of modes~\cite{brambila-multiplexed-spdc, onchip-lased-multiplexed-swfm-mahmudlu, zhang-multiplexed-nonpoled-spdc}.

Another desired property of photon pair sources is to produce narrowband photons because they are much easier to store in color center quantum memories. In the zero-added loss multiplexing (ZALM) protocol, quantum memories are used in the receiver stations between network endpoints to store the entangled qubit~\cite{ZALM, color-center-memory-network, shapiro-2024-ZALMsource, shapiro-2025-QDentanglementgen}. The ZALM protocol targets sources with tens to over a hundred frequency modes, photon bandwidths compatible with color center acceptance bandwidths (sub-GHz to tens of MHz), and free spectral ranges (FSR) large enough that individual modes are spectrally resolved by standard dense wavelength-division multiplexing channel spacings of approximately \SI{50}{\giga\hertz}. Our design targets of 106 modes, \SI{0.513}{\giga\hertz} bandwidth, and \SI{46.9}{\giga\hertz} FSR were chosen specifically to meet these requirements. Color centers are prominent candidates for quantum memories due to their long qubit storage times, but they have very narrow acceptance bandwidths that range from a few GHz to tens of MHz~\cite{nivc-doherty2013, sivc-bhaskar2020, tinvc-trusheim2020}. Cavity-based photon pair sources are of great interest for quantum networking, as they naturally produce narrowband photon pairs due to the Purcell enhancement of the density of frequency states around a cavity resonance~\cite{Guo2017, tsai-narrowband-spdc-for-qmemories, steiner-narrowband-cavity-spdc-bright, chen-narrowband-sfwm-integrated-cavity}. State-of-the-art cavity-based photon pair sources have produced sub-ten MHz bandwidth photons, but frequency multiplexed narrowband photon pair generation is still proving difficult~\cite{brambila-multiplexed-spdc, onchip-lased-multiplexed-swfm-mahmudlu, zhang-multiplexed-nonpoled-spdc, sfwm-multiplexed-algaas-pang}.

While color center quantum memories are promising candidates for qubit storage in quantum networks, they have only one wavelength acceptance band and do not natively support frequency-multiplexed operation. Additionally, the acceptance band of color centers is usually in the visible wavelength range instead of the telecom wavelengths used for fiber optic cables. This mismatch in operating wavelengths can be addressed by using sum frequency generation (SFG) to convert the telecom photons to the color center's acceptance band before loading the color center qubit. However, to determine the appropriate pump frequency to use for SFG, we need to measure the frequency state of the photon without destroying it. One way to generate an entangled photon pair whose frequencies are known is by performing a frequency-resolving heralding measurement on two entangled photon pairs. To perform this measurement, two photon pairs are required, both correlated in frequency and entangled in the qubit-carrying degree of freedom. Although frequency entanglement is not required for the ZALM protocol, it fulfills the requirement and enables other useful applications that make use of the extra degree of entanglement. Heralding also increases the success rate of distributed entanglement generation as it allows for higher average photon numbers to be sent while maintaining the same entangled state fidelity~\cite{ZALM}.

\begin{figure*}[th]
    \centering
    \includegraphics[width=0.9\linewidth]{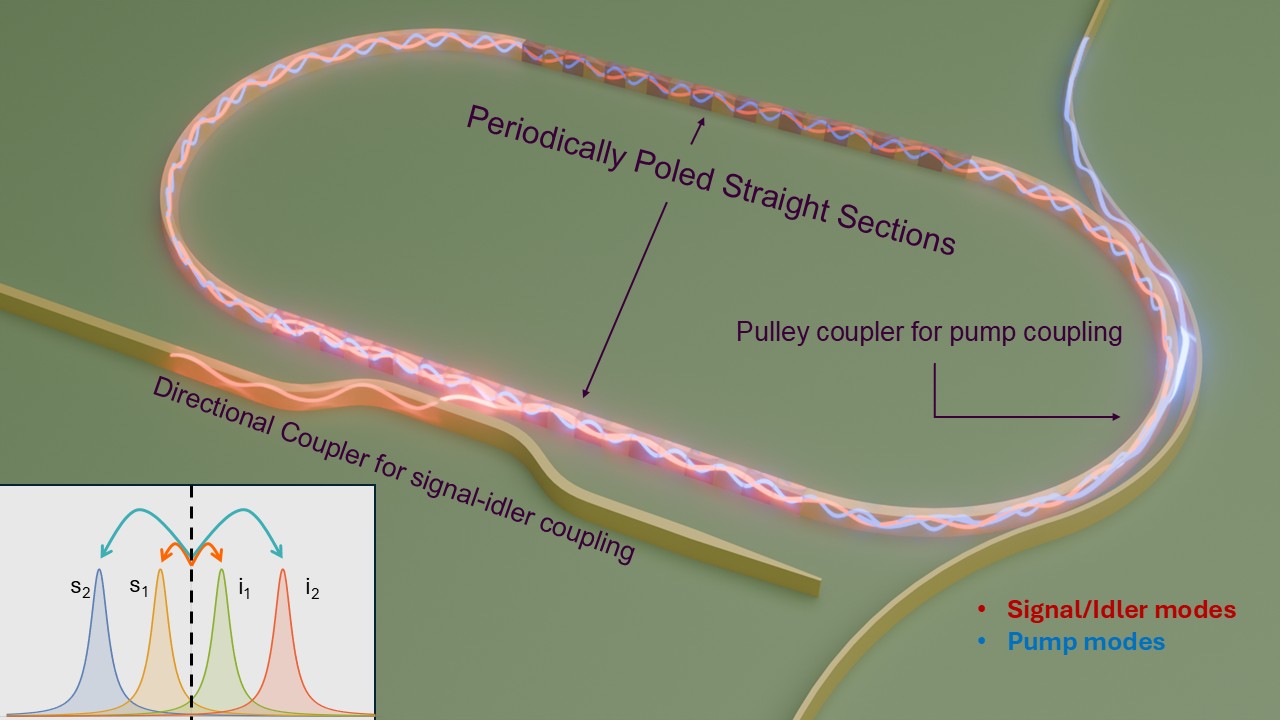}
    \caption{3D illustration of the proposed photonic microresonator structure with SPDC modes and their coupling in and out of the resonator highlighted.}\label{fig:Overview}
\end{figure*}

Fulfilling the complex requirements imposed by quantum networking protocols while ensuring the baseline performance of the source meets specifications is a significant engineering challenge. Many high-performance photon pair sources utilize nonlinear optical processes such as spontaneous parametric down-conversion (SPDC) or spontaneous four wave mixing (SFWM) to generate photon pairs, each with their own challenges. Very high PGR efficiencies---on the order of a few \unit{\giga\hertz\per\milli\watt}---have been achieved by SPDC sources due to high $\chi^{(2)}$ coefficients~\cite{slattery-cavity-spdc-review, efficient-ln-noncavity-source-tanzilli}, but the phase-matching bandwidth of cavity-based SPDC sources is quite small, usually only supporting a single resonance for periodically poled sources~\cite{Guo2017, brambila-multiplexed-spdc}, and showing poor PGR efficiency if periodic poling is omitted~\cite{zhang-multiplexed-nonpoled-spdc}. On the other hand, SFWM is often used for multiplexed sources as it is easier to quasi-phase match through dispersion engineering, which means the source waveguide does not need to be poled~\cite{chen-narrowband-sfwm-integrated-cavity, sfwm-comparison-multiplexed-kues}. However, the third-order nonlinear coefficient of most optical material platforms $\chi^{(3)}$ (approximately $10^{-20}$--$10^{-17}$~\unit{\square\meter\per\watt}) is several orders of magnitude smaller than the second-order coefficient $\chi^{(2)}$ (approximately $10^{-12}$~\unit{\meter\per\volt}), generally resulting in smaller PGR efficiencies by several orders of magnitude compared with SPDC when the same number of nonlinear interaction modes are resonant in both (i.e., both SPDC and SFWM are double resonant or triple resonant) and the same nonlinear material is used for both. One outlier to this pattern is found among sources on III--V platforms such as AlGaAs, which have very high third-order nonlinear coefficients ($n_2 = 1.6 \times 10^{-17}$~\unit{\square\meter\per\watt})~\cite{wang-photon-pair-source-review, baboux-moody-algaas-review-2023}. This makes achieving high PGR efficiency in SFWM processes possible, but the fabrication technology for III--V platforms is not as mature compared with materials such as silicon and Si\textsubscript{3}N\textsubscript{4}, making monolithic integration with other photonic devices challenging~\cite{steiner-narrowband-cavity-spdc-bright, sfwm-multiplexed-algaas-pang}.

In this paper, we demonstrate a material platform-agnostic simulation pipeline used to design an integrated photonic cavity-based entangled photon pair source that can produce frequency-entangled, frequency-multiplexed photon pairs. We take advantage of the high nonlinear susceptibility of SPDC, but by only poling a straight waveguide segment of a racetrack resonator, we enable our design to be compatible with x-cut and z-cut fabrication processes by making small adjustments to device geometry and waveguide poling period. We derive a formula for the JSA of our source in the large quality factor regime and use it to predict the joint spectral intensity and the pair generation rate efficiency of an example racetrack resonator source. Finally, we explore the design challenges and trade-offs inherent to designing a racetrack resonator photon pair source. We simulate an example source design on a thin-film lithium niobate (TFLN) platform, and predict a total PGR efficiency of \SI{1.75}{\giga\hertz\per\milli\watt} across 106 resonant mode pairs with average bandwidths of \SI{0.513}{\giga\hertz}---which is comparable to the scale of acceptance bandwidths of several color center quantum memories---and a free spectral range of \SI{46.9}{\giga\hertz}. Table~\ref{tab:sourceperf-comparison} compares the performance of some cavity-based entangled photon pair sources in the literature with the simulated performance of our source. The principal advantage of our design is the size of the supported frequency state space rather than exceptional per-island efficiency: the total PGR summed across all islands and the brightness per unit bandwidth are competitive with the best integrated TFLN results in the literature, while the effective frequency-space Schmidt number and per-island purity indicate a large, high-quality frequency-entangled state.

Section~\ref{sec:cavity-design} outlines the major design choices for our photon pair source design and outlines the simulation methods used to build the classical resonator model. In Sec.~\ref{sec:simulation}, we derive analytic formulas that can predict the quantum characteristics of our source based on classical cavity parameters. Finally, Sec.~\ref{sec:design-opt} explores some of the performance trade-offs based on the cavity's characteristics and design parameters.

\section{Racetrack Source Design}\label{sec:cavity-design}

The three main factors that determine the shape of the JSA for a cavity source are energy conservation, momentum conservation (phase matching), and resonance structure, as we derive in Appendixes~\ref{sec:AppA-straightwgJSAderiv} and \ref{sec:AppB-cavityJSAderiv}. Energy conservation is the easiest to engineer and design, since it is completely determined by the pump's frequency spectrum. We design for a pulsed Gaussian pump laser with a narrow bandwidth to eliminate off-diagonal islands in the low average photon pair regime.

Next, we use cavity resonances to constrain the JSA of photon pairs into densely spaced, narrow joint frequency modes (which we call ``islands''), which are desirable for producing high-fidelity frequency entanglement~\cite{ZALM,jsa-shape-theory-Jeronimo-Moreno2010}. Crucially, we want each island to have a factorizable form, which is required for single temporal-mode pair generation---in the Schmidt decomposition, this corresponds to high single-island purities~\cite{Raymer05-pure-state-generation}. For our source to generate dense, distinct islands, the signal and idler should both have narrow bandwidths, meaning the signal and idler modes both need to be resonant, which we will call double resonance, while we will call triple resonance the case when the signal, idler, and pump modes are all resonant. In order to maintain energy conservation in the degenerate opposite-sideband SPDC process, the pump frequency must be chosen such that the half-pump frequency point $\omega_p/2$ is placed either at a resonance or halfway between two signal and idler resonances. We choose to have the half-pump point be placed halfway between two resonances because it will allow us to split the generated photon pairs by wavelength without having an island where the signal and idler frequencies are the same~\cite{Reimer-2016-freqcomb, Gianini-2026-racetrack-SIsplit}. This process is still degenerate SPDC, but the cavity resonances together with energy conservation constrain the photons to be only in opposite resonant modes. Deterministic wavelength-based splitting changes the relevant state-space counting: islands mirrored about the $\omega_s=\omega_i$ diagonal are then treated as the same logical mode pair, so the accessible number of distinct islands is halved. The probability weight in the retained islands is correspondingly doubled, which must be accounted for when choosing the pump operating point. This opposite-sideband configuration is common in the literature because it provides a substantially broader phase-matching bandwidth than same-frequency signal and idler processes.

Requiring double resonance also lets us design denser island modes without causing mode overlaps, which increases the total entanglement-generation capacity of a network using our source. The downside is that constraining the pump frequency relative to the signal and idler resonances makes triple resonance very difficult in practice, so the nonresonant pump is the practical operating regime for this design. In Sec.~\ref{sec:design-opt}, we show that high-Q pump resonances give smaller PGRs than highly overcoupled cavities when the pump is maximally detuned from resonance. Therefore, under the non-resonant-pump constraint, we design the resonator to be highly overcoupled at the pump wavelength to maximize PGR. If triple resonance could be achieved through active tuning, the predicted PGR would increase by several orders of magnitude; we treat that as the theoretical ceiling of this design and leave it to future work.

The third effect---phase matching---can only be properly discussed once the cavity structure and the nonlinear process to produce the photon pairs have been chosen. The straightforward choice of structure for an integrated photonic cavity source is a ring or racetrack resonator as they are well studied, and photonic fabrication technology is mature, capable of producing resonators with quality factors between $10^6$ and $10^8$ which correspond to bandwidths on the order of 100 to 1~MHz at 1550~nm, which are on the order of commonly used color center acceptance linewidths~\cite{zhu-highQ-cavity-2021, sun-highQ-AlN-2019, ji-highQ-SiN-2021}. We choose to use SPDC instead of SFWM for our source design to achieve higher PGR efficiency because the $\chi^{(2)}$ coefficient is in general much larger than $\chi^{(3)}$. Since $\chi^{(2)}$ processes are only possible in materials with asymmetries in their crystal lattice, the refractive index of such materials depends in general on the polarization direction of the electric field, making the choice of propagation direction relative to the crystal axes important. Type-0 SPDC has been demonstrated before in ring resonators, mostly in z-cut materials where the SPDC waveguide modes do not change as the light changes propagation direction~\cite{Guo2017, ma-ppring-spdc-2020}. However, ring resonators do not scale well to larger sizes as they take up much more surface area on the chip than a racetrack cavity of the same length. Ring resonators also cannot be used for SPDC on an x-cut chip because the extraordinary axis would be in the wafer's plane, making periodic poling very difficult if not impossible, as the light propagates partially along the extraordinary axis in a ring resonator. Therefore, following the approach of several published works on integrated racetrack sources~\cite{mckenna2022ultralow, park2024squeezed,hwang2024spdc}, we choose a racetrack configuration with only one straight waveguide section periodically poled, as shown in Fig.~\ref{fig:Overview}, to more efficiently use wafer surface area and to ensure compatibility with x-cut chips. Furthermore, the length of the straight waveguide sections can be easily changed in the design process to adjust the resonator's FSR, PGR efficiency, and quality factor. By only periodically poling a straight section of the racetrack, we can support x-cut chips as the light can be polarized along the extraordinary axis throughout the length of the poled section. On the waveguide bends, the SPDC and the reciprocal SFG processes are not phase matched, meaning the power in the signal and idler modes stays constant as they traverse the bends except for a small sinusoidal oscillation. We note that for a z-cut wafer higher power conversion efficiency could be achieved by poling the waveguide bends as well, but our configuration does work for a z-cut wafer.

We note that one downside of our design is that it is difficult to use with type 2 SPDC because both signal and idler modes need to be resonant at the same frequencies. For nonisotropic materials, light polarized along two different directions will experience different refractive indices. This will result in an FSR dispersion between the two wavelengths, making it very difficult to align the cavity resonances. This will cause the line of doubly-resonant islands to not be collinear with the energy-conservation envelope, which will significantly limit the number of islands supported. Type-1 SPDC is supported by our source, but it has a smaller nonlinear susceptibility coefficient for most materials than type-0 SPDC, making it worse for our use case.

While SPDC has high nonlinear coefficients, quasi-phase matching (QPM) by periodic poling usually results in a narrow phase-matching bandwidth due to refractive index dispersion causing a bigger phase mismatch the further we are from the ideal phase-matched frequency. Material dispersion differs significantly between usual pump (visible) and signal and idler (telecom) wavelengths, which leads to a large phase mismatch for down-conversion processes when off-center from the ideal QPM frequency. For straight waveguide sources, this effect restricts SPDC to a very narrow bandwidth because the source is usually made to be very long (on the order of centimeters) to achieve high power conversion efficiency. We next explain why this principle will be true for our cavity source as well. Cavity modes maintain a constant phase pattern vs distance traveled in steady state both on- and off-resonance, with the on-resonance case being trivial. Off-resonance, the phase mismatch between round trips causes destructive interference, which in turn forces any light that does not match the previous trip's phase to couple out of the cavity. This results in a weaker amplitude but a stable phase pattern for nonresonant light traveling in the resonator. Therefore, the relative phase for each SPDC mode at the starting point of the periodically poled waveguide is the same over all round trips, which means our source can be treated as a periodically poled straight waveguide inside an optical cavity for phase matching purposes. Thus, we can adjust the phase-matching bandwidth of the source by changing the length of the periodically poled straight waveguide section.

\begin{figure}[t]
    \centering
    \includegraphics[width=1\linewidth]{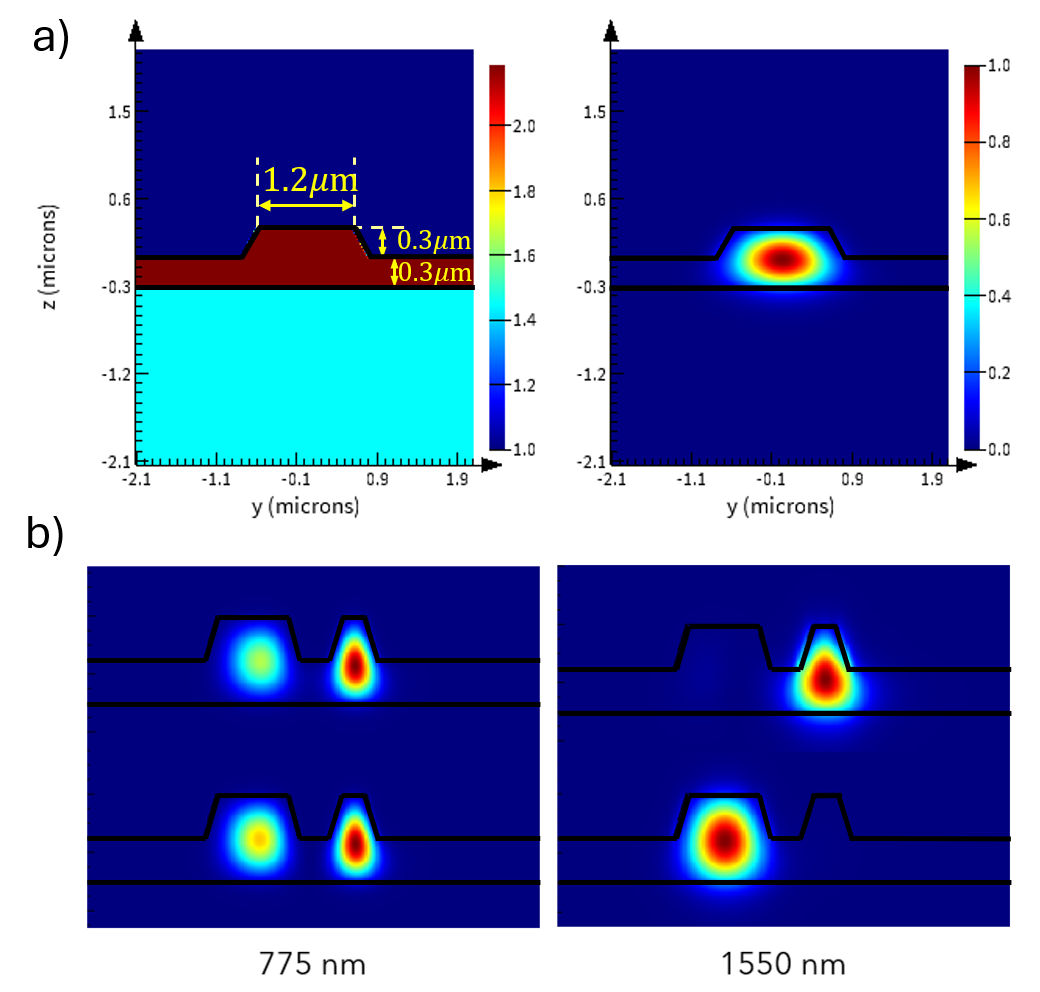}
    \caption{(a) Racetrack waveguide geometry and fundamental TE mode at 1550 nm plotted with 1:1 aspect ratio. (b) Even and odd supermodes of the pulley coupler. Negligible overlaps between the modes at 1550 nm indicate very low coupling between the racetrack and pulley waveguides. The modes were solved for at the apex of the pulley coupler bend.}\label{fig:pulley-coupling}
\end{figure}

A significant design target for a photon pair source is high PGR efficiency, which is proportional to the square of the periodically poled waveguide section's length. A consequence of making a periodically poled waveguide source longer is a narrower phase-matching bandwidth. Generating narrow photon pairs by engineering a narrow phase-matching bandwidth is a common design target for photon pair sources, but it is undesirable for our case. Instead, we want a broad phase-matching bandwidth to support more islands, which are then narrowed down by cavity resonances. Our source's phase-matching bandwidth depends inversely on the length of the periodically poled straight section, which means there is a direct trade-off between the number of islands supported and PGR efficiency. Fortunately, we can compensate for this trade-off by generating denser islands without changing the length of the periodically poled section. We can do this by lengthening the straight waveguide sections while keeping the length of the poled waveguide section the same. This will decrease cavity FSR and make the islands more densely spaced. However, as we will discuss in Sec.~\ref{sec:design-opt}, there is an upper limit to how dense we can space the cavity resonances that is determined by the quality factor. The islands must be separated well enough to produce Schmidt modes that are well-separated in order to produce island modes that collectively form a high-fidelity frequency-entangled state. This requirement can be satisfied as long as the island bandwidths are much smaller than the FSR of the cavity.

\begin{figure*}[!ht]
    \centering
    \includegraphics[width=1\linewidth]{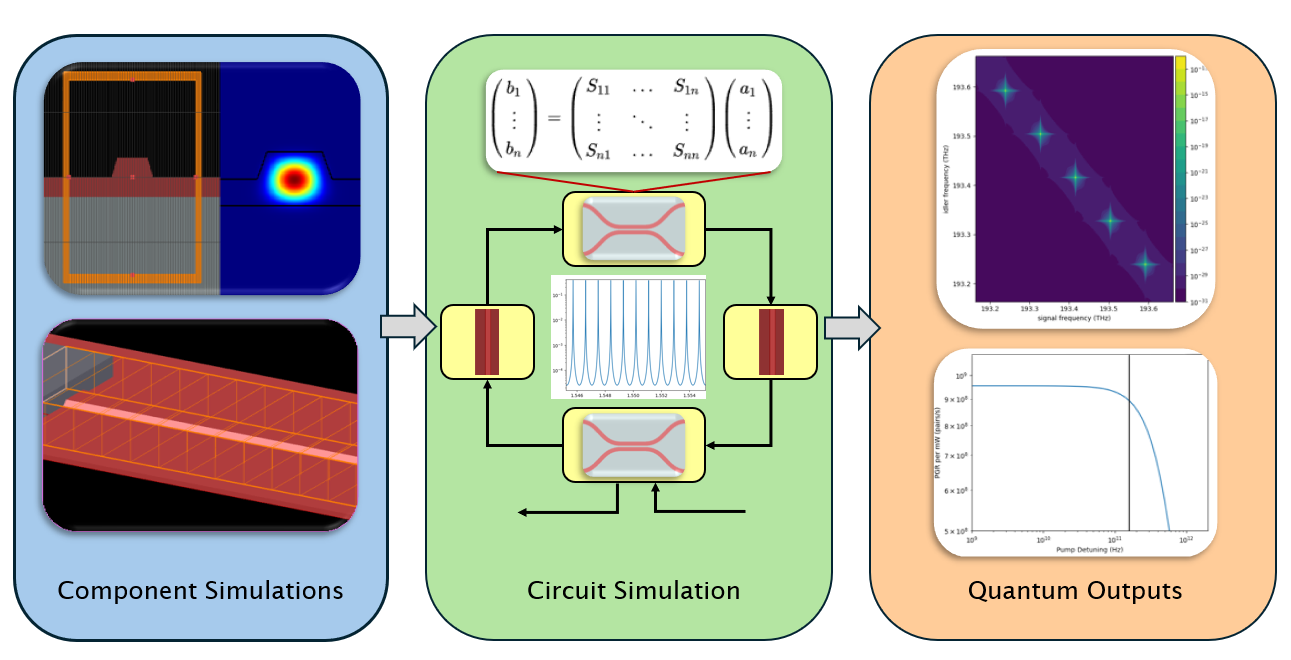}
    \caption{Overview of the simulation pipeline used to generate the numerical quantum-state outputs. Component simulations produce scattering-parameter matrices that feed the circuit simulator, which returns the cavity spectrum. Those extracted cavity parameters are then used to calculate the predicted JSA and PGR.}\label{fig:simulation-pipeline}
\end{figure*}

To meet the desired narrow bandwidth targets and the above mentioned design goals, our resonator must have a high quality factor in the signal and idler frequency range, but the pump should be highly overcoupled to the resonator. This is difficult to achieve with only standard directional couplers due to the high variance of directional couplers' coupling constants vs frequency. Instead, we use a pulley coupler to inject the pump light into the resonator and achieve high coupling at the pump frequency without coupling out signal and idler photons, similar to Guo \textit{et al.}~\cite{Guo2017}. By making the pulley waveguide's cross section subsingle mode at telecom wavelengths but single-mode at visible wavelengths, we can inject pump light into the resonator without causing significant loss in the signal and idler pairs. The gap in the pulley coupler is optimized for maximum supermode overlap at the pump wavelength while the overlap for the telecom modes is very small, as illustrated by Fig.~\ref{fig:pulley-coupling}, thus optimizing for selective pump coupling. We can then use a directional coupler on one of the straight waveguide sections to couple telecom light in and out of the resonator and extract the produced photon pairs. By adjusting the coupling gap and length, we can optimize for high telecom coupling but low visible coupling. A very high coupling coefficient for the directional couplers will reduce the quality factor of the resonator, which is generally not desirable, but we do want a coupling coefficient high enough that most of the produced photon pairs are coupled to the bus waveguides instead of lost due to radiative losses from the resonator. We explore this design trade-off further in Sec.~\ref{sec:design-opt}.

Having made the above design choices, we now choose specific geometry parameters and a material platform to simulate and predict how an actual source would behave when measured experimentally. We design a racetrack source on an x-cut TFLN on SiO\textsubscript{2} platform with \SI{600}{\nano\meter} LN thickness, \SI{300}{\nano\meter} etch depth, and \ang{55} sidewall angle, a layer stack similar to ones used for electro-optic modulator systems. We choose a large ring waveguide top width of \SI{1.2}{\micro\meter} to achieve lower propagation losses, a \SI{0.4}{\micro\meter} pulley waveguide top width to make the pulley waveguide single-mode around \SI{775}{\nano\meter} and have no guided modes at \SI{1550}{\nano\meter}, and a \SI{0.52}{\micro\meter} coupling gap measured from the base of the waveguide. We calculate the pulley coupler's single-pass power cross-coupling coefficient to be 21\% in the visible range while having less than 1\% loss in the telecom range.

\section{Simulating Cavity Source Quantum Properties}\label{sec:simulation}

\begin{figure}[htbp]
    \centering
    \includegraphics[width=0.9\linewidth]{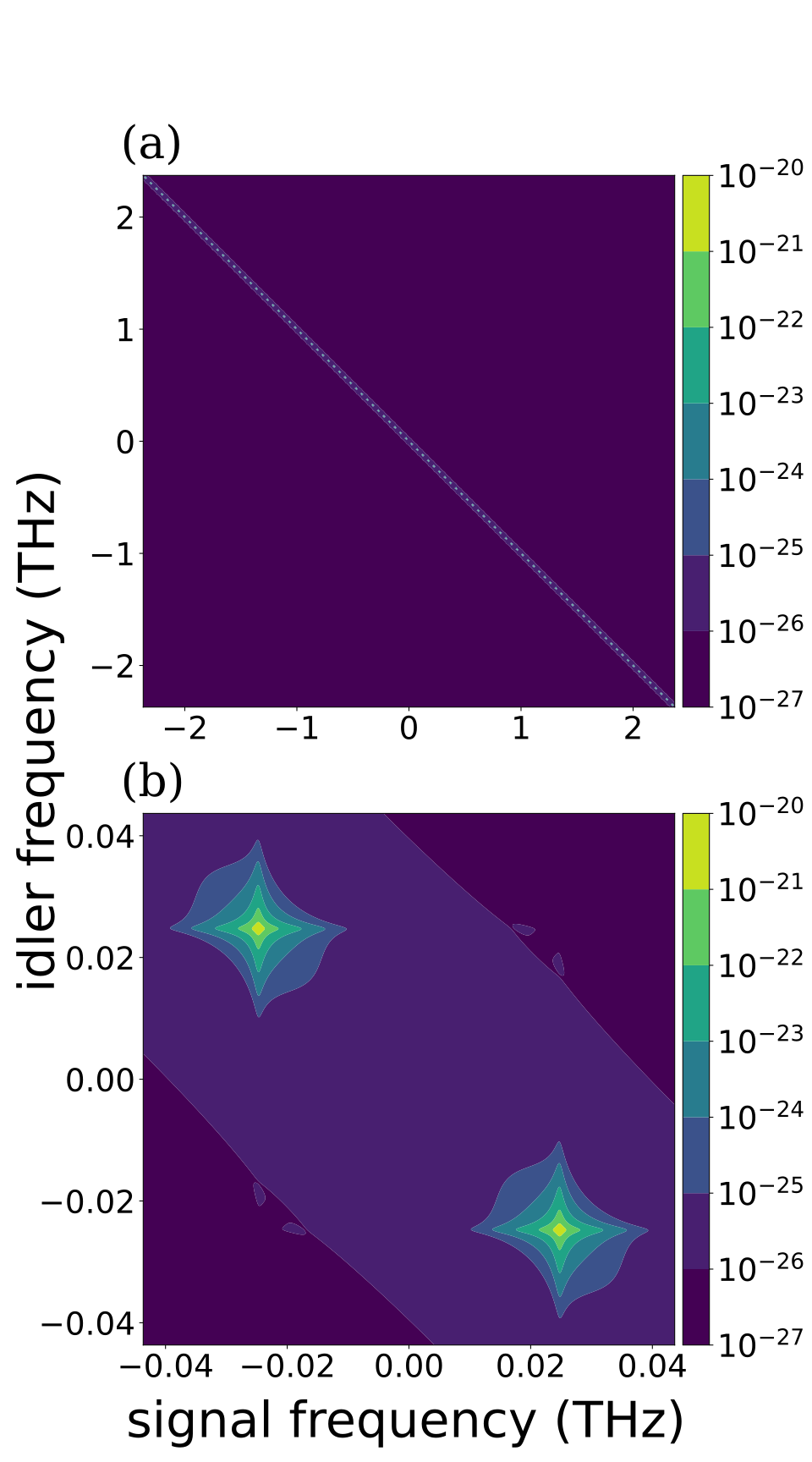}
    \caption{Contour plots of the JSA function for a cavity SPDC source (a) and an enlargement of one resonant island (b).}\label{fig:jsa-simple}
\end{figure}

Both classical and quantum electromagnetic theory can be based on the same mode functions, sometimes called the classical-quantum correspondence principle. Therefore, we can predict the JSA and PGR of our racetrack resonator by solving for the cavity frequency spectrum and scattering parameters using classical simulations and substituting the results into analytic quantum theory.

Figure~\ref{fig:simulation-pipeline} illustrates the three main steps of the simulation process: component simulations, circuit simulations, and the quantum properties that are calculated. First, we simulate components with various electrodynamics simulation methods: Lumerical finite difference eigenmode solver for waveguide mode solving, Lumerical finite difference time domain solver (FDTD) for the waveguide bends and the pulley coupler at telecom wavelengths, and the open-source FDTD solver Meep~\cite{Oskooi2010} for the directional couplers. Because of the large sidewall slant and shallow etch of the chosen layer stack, the pulley coupler and waveguide bends are too large to simulate in the visible wavelength range using 3D FDTD methods, and 2D FDTD does not consider the effect of sidewall angles and anisotropy. We can get around this limitation by using a segmented straight waveguide in Lumerical eigenmode expansion solver (EME) to approximate the bend and pulley scattering parameters in the pump wavelength range. We select cross-sectional slices from the bent waveguides at many points, calculate the refractive index of the waveguide material along the TE-polarized direction, and set that as the refractive index data for a short straight waveguide. By connecting several of these segments, we can use EME to approximate the scattering parameters of a bent waveguide structure. We verified that this method approximates the scattering parameter magnitudes to within a few percent error by comparing the output of our sectioned EME method with FDTD simulations of waveguide bends at telecom wavelengths where the FDTD simulation of the bend is not as computationally intensive. However, this method introduces significant phase error in the scattering parameters, which means we cannot use this method to predict the exact location of resonances in the visible wavelength range, but the pump quality factor (and consequently the cavity decay rate) can be closely approximated using this method. More details about our segmented EME method can be found in Appendix~\ref{sec:AppD-segmented-EME}.

The scattering parameter matrices extracted from the finite difference simulations can be combined in a photonic circuit solver such as Lumerical Interconnect or the Python package \texttt{sax}~\cite{sax}. The circuit solver gives us the frequency spectrum of the resonator, which we then fit to a linear combination of Lorentzian functions, extracting the cavity's resonant frequencies, total cavity decay rates for each resonance, and coupling constants. The final part of our simulation pipeline is a theoretical model that can predict the joint spectral amplitude and pair generation rate functions from the resonance parameters of our source.

\begin{figure*}[!t]
    \centering
    \includegraphics[width=1\linewidth]{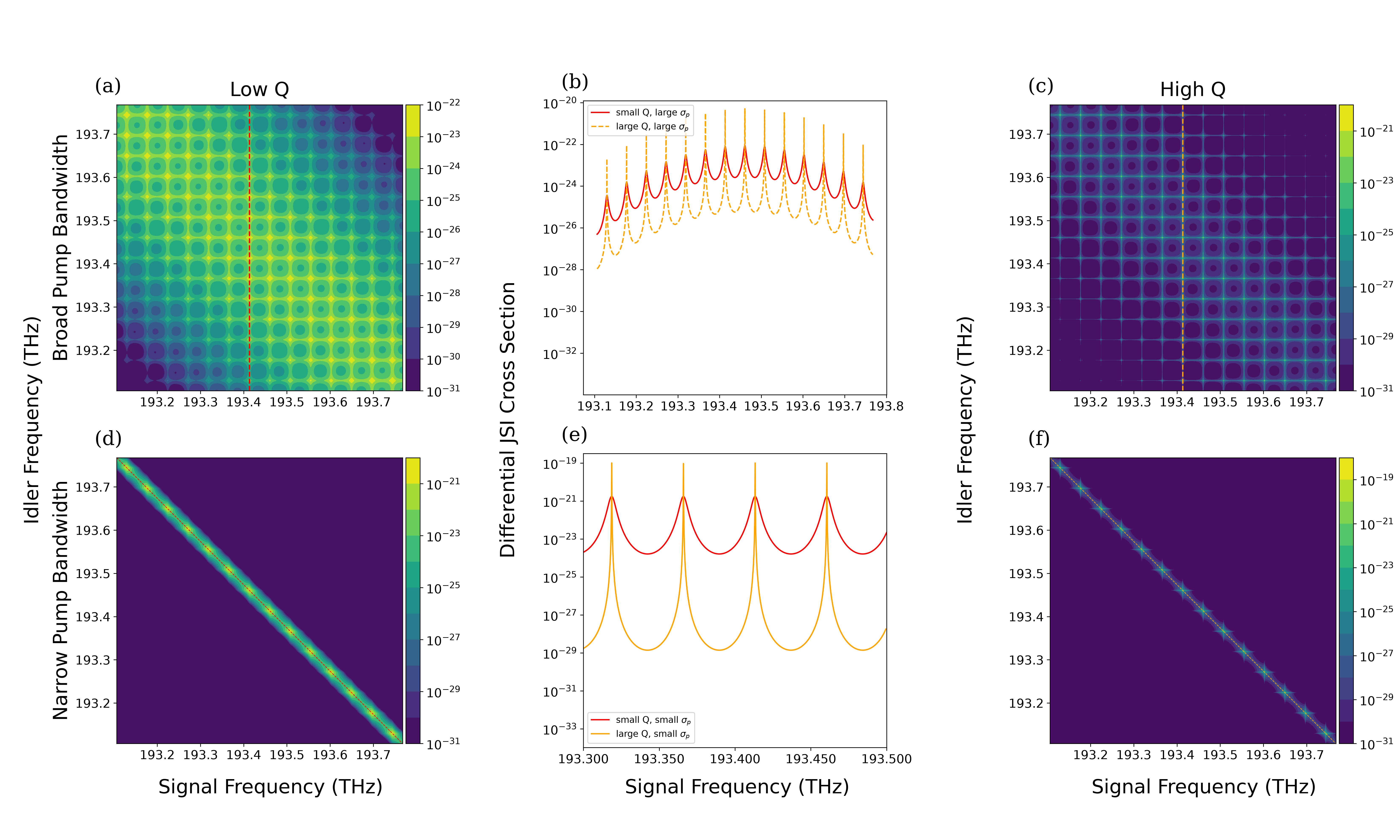}
    \caption{Comparison of various resonant cavity SPDC source joint spectral intensities. In the contour plots, the rows correspond to a broadband pump of $\sigma_{f} = \SI{100}{\giga\hertz}$ [(a)--(c)] and a narrowband pump [(d)--(f)] of $\sigma_{f} = \SI{5}{\giga\hertz}$, where $\sigma_{f}$ is the RMS width of the pump power spectrum, while the columns represent low Qs of $Q_{\text{low}} \approx 4 \times 10^4$ [(a),(d)] and high Qs of $Q_{\text{high}} \approx 2 \times 10^6$ [(c),(f)]. The middle plots show specific cross sections of the JSI plots, enlarged to show detail. Plot (b) shows a vertical cross section (constant $\omega_s$) for the broad pump regime, demonstrating how the frequency state has undesirable side islands, with probability densities comparable to that of the main island. Since the JSA formula is not normalized, the function values are proportional to the cavity PGR. Therefore, the peaks extending past the low-Q cross section illustrate the cavity enhancement of PGR at resonance. Plot (e) shows the cross sections along the $\omega_p = \omega_s + \omega_i$ diagonal from plots (d), (f) and illustrates how higher Q creates narrower islands.}\label{fig:jsa-comparison}
\end{figure*}

We now introduce a derivation of the joint spectral amplitude and pair generation rates from our cavity source. We only list the main ideas and formulas here; the full derivation is detailed in Appendixes~\ref{sec:AppA-straightwgJSAderiv}--\ref{sec:AppC-cavity-enhancement}. We start with the interaction Hamiltonian
\begin{equation}\begin{split}
H^{(2)} = \sum_{j,k} & \int_{-\infty}^{\infty}\int_{-\infty}^{\infty}\int_{-\infty}^{\infty} \frac{d\omega}{2\pi} \frac{d\omega'}{2\pi} \frac{d\omega_p}{2\pi} e^{-i\Delta \omega t}\\
\times & G_{jk}(\omega, \omega', \omega_p) \hat{b}_j^\dagger(\omega) \hat{b}_j^\dagger(\omega') \hat{b}_k(\omega_p) + \text{H.c.}\text{,}
\end{split}\end{equation}
where $G$ is the nonlinear interaction coefficient between waveguide modes $j$ and $k$, given by
\begin{widetext}\begin{equation}
    G_{jk}(\omega,\omega', \omega_p) = i \frac{\sqrt{2}\ \hbar^{\frac{3}{2}}L}{\pi\sqrt{\varepsilon_0}}   \sqrt{\frac{\omega\ \omega'\ (\omega_p)}{v_{g,j}(\omega) v_{g,j}(\omega') v_{g,k}(\omega_p) }} O_{jk}\left(\beta_j(\omega), \beta_j(\omega'), \beta_k(\omega_p)\right)\ \frac{\chi^{(2)}\left(\omega_p; \omega, \omega' \right)}{n_j^2(\omega) n_j^2(\omega') n_k^2(\omega_p)} \text{sinc}\left(\frac{\Delta \beta\ L}{2}\right)\text{,}
\end{equation}\end{widetext}
where the functions $O\text{,}\ F^{(2)}\text{,}\ \Gamma\text{, and}\  \Delta\beta$ are the triple mode overlap integral, nonlinear interaction coefficient, pump frequency spectrum, and wave vector mismatch functions, respectively. The full definition of each constant is listed in Appendix~\ref{sec:AppA-straightwgJSAderiv}. After assuming each mode is in the fundamental TE mode, we can substitute the Hamiltonian into the Heisenberg equation for a straight waveguide mode's annihilation operator $ \hat{b}(\omega)$ to its evolution between the input and output of the waveguide.
\begin{equation}\begin{split}
    \partial_t \hat{b}(\omega) &= \frac{i}{\hbar}[H^{(2)}, \hat{b}(\omega)]\\
    &\begin{split}
    = \frac{i}{\hbar} \int_{-\infty}^{\infty} \int_{-\infty}^{\infty} &\frac{d\omega_p}{2\pi}   \frac{d\omega'}{2\pi} G(\omega, \omega', \omega_p)\\
    & e^{-i\Delta \omega t}\ \alpha_p(\omega_p)\ \hat{b}^\dagger(\omega')\text{.}
    \end{split}
\end{split}\end{equation}
Integrating this equation over the propagation time through the waveguide for the output operator gives us
\begin{equation}\begin{split}
    \hat{b}_{\text{out}}(\omega) \approx \hat{b}_{\text{in}}(\omega) + &\\
    \int_{0}^{\infty} \frac{d\omega'}{2\pi} &J(\omega,\omega') \hat{b}_{\text{in}}^\dagger(\omega')\text{,}
\end{split}\end{equation}
where
\begin{equation}
J(\omega,\omega') = \frac{i}{\hbar} G(\omega,\omega', \omega+\omega')\ \alpha_p(\omega + \omega')\text{,}
\end{equation}
and $\alpha_p(\omega + \omega') $ is the coherent-state amplitude for the pump, which is straightforward to calculate given the pump pulse energy, namely $\alpha_p (\omega + \omega') = \sqrt{\braket{\hat{n}_p}} = \sqrt{ \int dt P_p(t)\ \Gamma(\omega + \omega')  / (\hbar\ (\omega + \omega'))}$ where $P(t)$ is the pump pulse's power over time and $\Gamma(\omega_p)$ is the frequency spectrum of the pump pulse normalized such that $\int d \omega_p\ \Gamma(\omega_p) = 1$. The function $J(\omega,\omega')$ can be shown to effectively be a joint spectral amplitude function for a straight waveguide source. We can then use the cavity boundary conditions
\begin{align}
\hat{c}(0^+,\omega) &= \sigma \hat{c}(L^-,\omega) + i\kappa \hat{a}_{in}(\omega)\text{,} \\
\hat{a}_{out}(\omega) &= i\kappa \hat{c}(L^-,\omega) + \sigma \hat{a}_{in}(\omega)\text{,}
\end{align}
where $\sigma$ is the cavity reflection coefficient and $\kappa$ is the cavity transmission coefficient for signal and idler wavelengths, to write down the evolution of the output mode annihilation operator in terms of the input ladder operators,
\begin{equation}\begin{split}
    \hat{a}_{out}(\omega) = h(\omega)&\hat{a}_{in}(\omega) +\\
    &e^{i\beta(\omega)L} \int_0^\infty \frac{d\omega'}{2\pi} \, j(\omega,\omega') \hat{a}_{in}^\dagger(\omega')\text{,}
\end{split}\end{equation}
where the functions $h$, $j$ are given by
\begin{align}
h(\omega)  &= -e^{i\beta(\omega)L} \frac{\eta - \sigma e^{-i\beta(\omega)L}}{1-\sigma\eta e^{i\beta(\omega)L}}\text{,}\\
j(\omega,\omega') &= \frac{\kappa^2 \eta J(\omega,\omega')}{(1-\sigma\eta e^{i\beta(\omega)L})(1-\sigma\eta e^{-i\beta(\omega')L})}\text{.} \label{eq:JSA}
\end{align}

The function $j(\omega,\omega')$ is the joint spectral amplitude function of the cavity source when measured right outside the cavity, and it is equal to the straight waveguide source's JSA multiplied by the cavity density of states for the two produced photons. The joint spectral intensity (JSI) for our example design is plotted in Fig.~\ref{fig:jsa-simple}. The island structure is visible, and the double Lorentzian shape of each island indicates no intraisland correlations between signal and idler photons, meaning the photon pair is separable in frequency and its purity is high~\cite{Raymer05-pure-state-generation}. The temporal modes corresponding to these islands will therefore be decaying exponential functions.

Finally, we derive the pair generation rate for one island of the source. In the low average photon number regime, we can approximate the number of photon pairs produced using the expectation value of the number operator
\begin{equation}
    dR_{\text{pair}}(\omega) \approx \frac{1}{2} \braket{\hat{n}(\omega)} = \frac{1}{2} \braket{0_{in}| \hat{a}^\dagger_{out}(\omega) \hat{a}_{out}(\omega) |0_{in}}\text{.}
\end{equation}
We evaluate this expression to
\begin{equation}
    \braket{\hat{n}(\omega)} = \int_0^\infty \frac{d\omega'}{2\pi} |j(\omega,\omega')|^2\text{,}
\end{equation}
which is simply the JSA of the photon pair integrated across one of the frequency variables. Since the photon pairs are degenerate in frequency, this integral is simply the marginal spectral density of one of the photons. This indicates that the probability of generating a photon pair per island of the JSA is
\begin{equation}\label{eq:PGR}
P_{\text{island}} = \int_{-\Omega_{FSR}/2}^{\Omega_{FSR}/2}  \int_{-\Omega_{FSR}/2}^{\Omega_{FSR}/2} \frac{d\omega}{2\pi} \frac{d\omega'}{2\pi} \frac{\left|j(\omega,\omega')\right|^2}{2}\text{.}
\end{equation}
We use Eqs.~(\ref{eq:JSA}) and (\ref{eq:PGR}) to calculate the predicted JSA and PGR of our example racetrack resonator design and explore the design space of our resonator source.

\section{Simulation Results and Design Trade-offs}\label{sec:design-opt}

\begin{figure}[!htbp]
    \centering
    \includegraphics[width=1\linewidth]{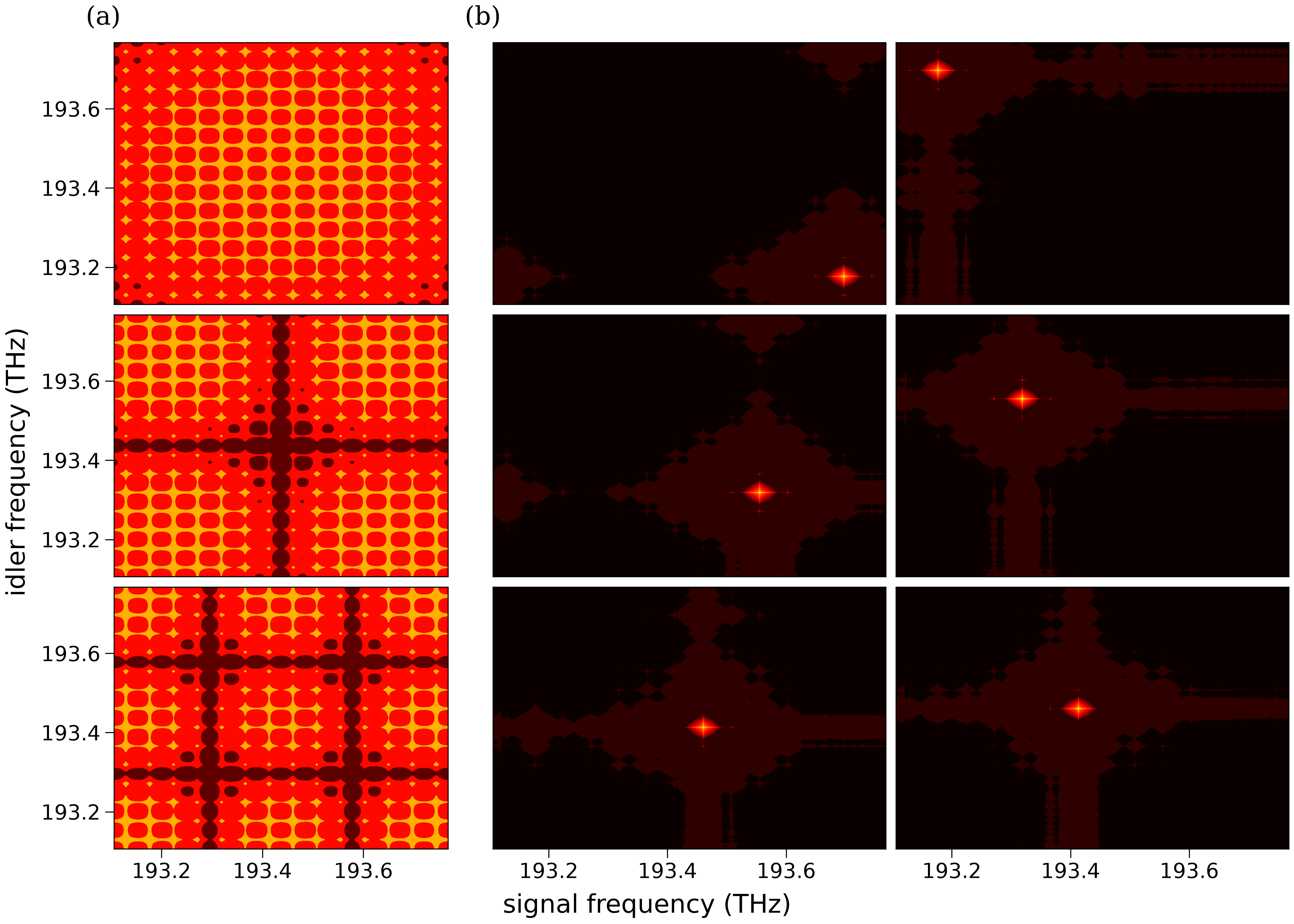}
    \caption{Comparison of Schmidt modes for a narrowband and a broadband pump. Plots (a),(b) are the decompositions of the JSA in plots (a),(f) from Fig.~\ref{fig:jsa-comparison}. Each panel shows the magnitude of one Schmidt mode, $|u_k(\omega_s)\,v_k(\omega_i)|$, normalized to its maximum. The distinct, minimally mixed Schmidt modes in the narrowband case illustrate the high quality of the frequency-bin entanglement.}\label{fig:schmidt-modes}
\end{figure}

We used our simulation pipeline and cavity model to explore how resonator parameter changes affect the output quantum state, and optimized our example source design for frequency-entangled, frequency-multiplexed photon pair generation at a high PGR. Figure~\ref{fig:jsa-comparison} shows the JSI for varying cavity quality factors and pump bandwidths. As explained in Sec.~\ref{sec:cavity-design}, the key functions and their corresponding physical parameters that determine the overall shape of the JSA are the pump frequency spectrum, the cavity transmission spectrum, and the phase matching function (periodic poling and waveguide dispersion). Each of these functions has a unique effect and purpose in engineering the JSA shape. The pump envelope function is set to be narrower than the phase matching bandwidth along the $\omega_s=\omega_i$ diagonal, which means that the pump bandwidth determines the thickness of the overall envelope of the JSA. The QPM function then defines the length of the JSA envelope along the $\omega_s + \omega_i = \omega_p$ diagonal, which determines the number of resonances that fit into the JSA envelope along the diagonal. We note that in the case the pump is broader than the QPM function, the QPM function determines the JSA envelope shape in all dimensions. Figure~\ref{fig:jsa-comparison}(c) shows that the JSA exhibits off-diagonal resonances when the pump is broadband, reducing the quality of frequency entanglement. Therefore, using a narrow pump bandwidth eliminates side islands that decrease fidelity with a perfectly frequency-entangled state. The effect of quality factor is easier to analyze, as a higher quality factor simply concentrates the frequency state probability at the resonances, narrowing the bandwidth of the produced photon pairs. We note that our high-Q approximation used in the JSA derivation means only the right figures are good numerical approximations, but our model still gives a qualitative estimate for the JSA in the low-Q regime. 

Once we have the JSA of our simulated circuit, we can calculate its Schmidt decomposition, which lets us investigate the quality of frequency entanglement in the produced photon pairs. We calculate the singular value decomposition of the 2D JSA array to calculate its singular values and arrays, which correspond to the Schmidt coefficients and modes. Figure~\ref{fig:schmidt-modes} shows an example in which we compare the Schmidt decomposition of a JSA produced by a cavity source with a narrowband pump and by a broadband pump. We limited the frequency span in these simulations, and we used the low-Q JSA of Fig.~\ref{fig:jsa-comparison}(a) for the broadband pump and the high-Q JSA of Fig.~\ref{fig:jsa-comparison}(f) for the narrowband pump, making the relative position of the various islands clearly visible. It is clear from the Schmidt decomposition that a pump that has a narrow bandwidth compared with the cavity FSR is desirable if the goal is to produce a high-fidelity frequency-entangled state, as it produces a JSA whose Schmidt modes are independent and do not mix between resonances. However, in order to ensure high joint spectral purity, the pump bandwidth must be broad enough to not truncate the double star shape. We note that the star shape (or a circular or elliptical shape) is not required for the individual islands' purity to be high. As long as the island shape is factorizable into a product of the two photons' marginal distributions, very large single-island purities can be achieved. The pump bandwidth truncating the edges of the double-Lorentzian stars truncates the JSA, making the joint state not a perfect product of two Lorentzians, causing the imperfect single-island purity~\cite{Raymer05-pure-state-generation}.

We evaluate the Schmidt decomposition numerically over the full JSA frequency interval using a grid that resolves the individual spectral islands. The same frequency resolution is used when restricting the calculation to individual islands to determine their spectral purities. Thus, the full-JSA Schmidt number and the single-island purities characterize, respectively, the global frequency-entangled state and the internal separability of its constituent islands.

\begin{figure}[!t]
    \centering
    \includegraphics[width=1\linewidth]{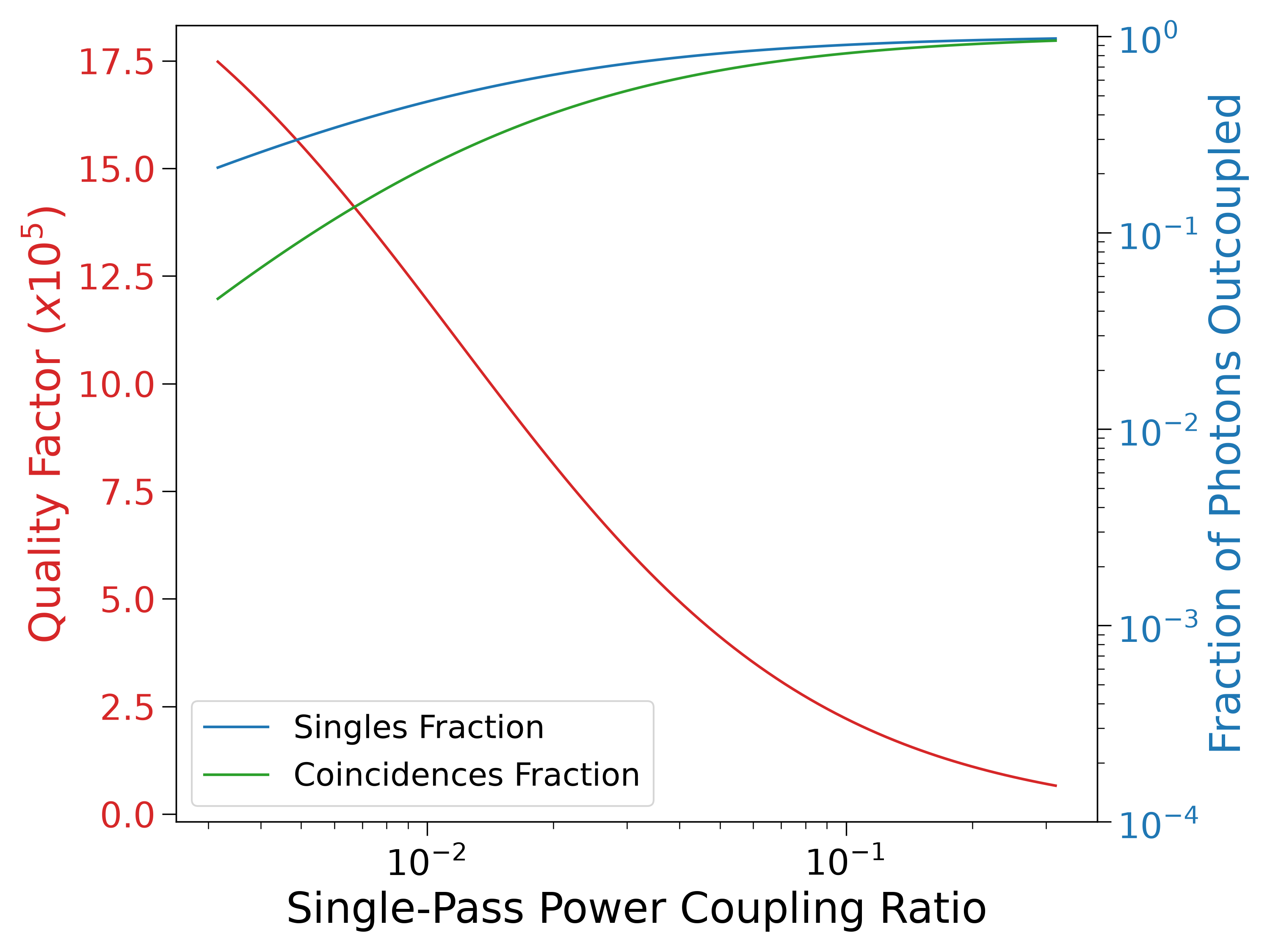}
    \caption{Cavity quality factor and the fraction of photons out-coupled successfully as a function of directional coupler single pass power cross coupling.}\label{fig:DC-coupling-tradeoff}
\end{figure}

\begin{figure}[!t]
    \centering
    \includegraphics[width=1\linewidth]{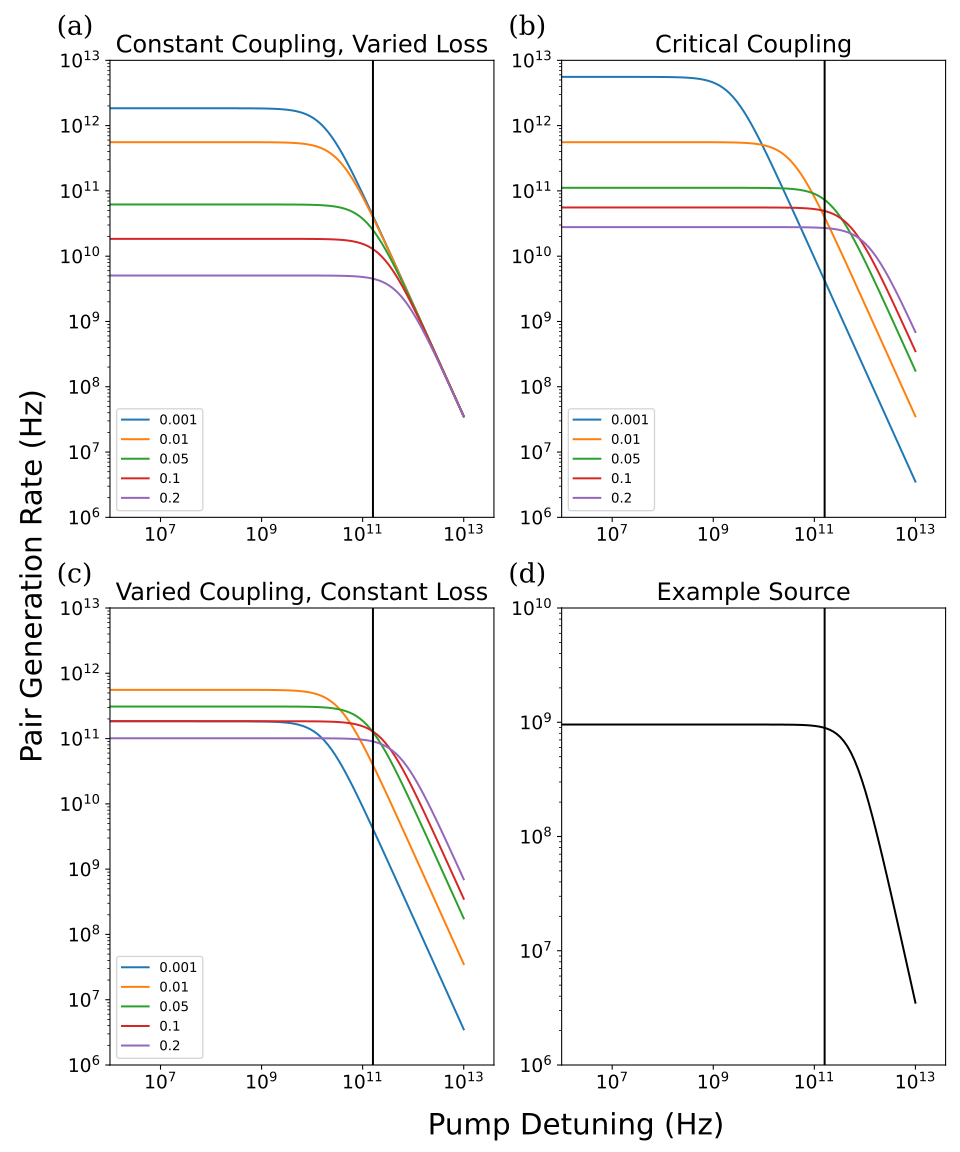}
    \caption{Simulated single-resonance pair PGRs as a function of detuning from resonance at the pump frequency for several single-pass effective coupling ratios for the input (pulley) coupler and a radiative loss channel. Plots (a)--(c) are simulations of semi-ideal devices with much smaller losses than a real device and are included to illustrate the design trade-offs. The numbers in the legend are the single-pass coupling ratio of the pump's input coupling and ``effective'' loss coupling channels. For the plots where one of the couplings has a constant value, the single-pass coupling ratio is fixed at 0.01. Note that the orange line is the same for plots (a)--(c) as all coupling ratios are set to 0.01 for those cases. All the plots except the bottom right plot have the same scale for easier comparison. Plot (d) is the actual device simulated with our pipeline but using the methods of Guo \textit{et al.} and corresponds to a pump Q of $2 \times 10^3$~\cite{Guo2017}. A vertical black line indicates the maximum possible detuning for all resonators which is at the halfway point between two resonances in the \SI{775}{\nano\meter} range.}\label{fig:pgr-comparison}
\end{figure}

We have established that a high Q-factor is desirable because it results in narrower photon bandwidths and increased PGR efficiency. To achieve high Q, a resonator must have low radiative losses and small coupling coefficients. In the photonic cavity design process, we try to minimize radiative losses such as waveguide propagation loss and mode mismatch losses between straight and bent waveguides, but there is always a baseline loss in real photonic resonators based on the waveguide geometry constraints set by the foundry's layer stack properties, intrinsic fabrication variability, waveguide surface defects, and in our case the pulley coupler. These effects restrict real resonators to a maximum Q determined by these losses $Q \leq \omega_0/2 R_{loss}$ for a cavity decay rate due to radiative loss $R_{loss}$, assuming optimal waveguide and layout geometries. If we want to approximately maximize cavity Q in our source, we can optimize the directional couplers for very small coupling. However, exceedingly low coupling will result in a large portion of the produced photon pairs being lost to waveguide propagation losses instead of being coupled to the bus waveguides and extracted from the source. Therefore, there is a direct trade-off between the effective PGR and the Q of our source, as shown by Fig.~\ref{fig:DC-coupling-tradeoff}. This trade-off must be considered when adapting our source to specific quantum memories since most memory implementations (e.g., cold atoms, vacancy centers) have narrow acceptance bandwidths. Designing the cavity for high quality factors by adjusting the cavity coupling will reduce the ratio of photon pairs successfully outcoupled, effectively reducing the pair generation rate. Our example design uses a directional coupler with a single-pass power cross-coupling coefficient of $\kappa^2 = 0.052$, which results in a Q of about $3.78 \times 10^5$ for the simulated resonators.

The PGR efficiency of our source is significantly affected by the pump's resonance properties and the choice of pump coupling constant is not straightforward, as illustrated by Fig.~\ref{fig:pgr-comparison}. While our cavity source model does not support a resonant pump, we can use the methods in Guo \textit{et al.} for calculating the pair generation rate of a degenerate cavity SPDC source to explore the effects of pump resonance~\cite{Guo2017}. Figure~\ref{fig:pgr-comparison}(a) shows how increased radiative losses result in a lower on-resonance PGR, but do not affect off-resonance PGR, since the amount of power lost is proportional to the power itself. Therefore, the effect of radiative losses is highest on-resonance, where the pump field amplitude is amplified in the resonator. We predict that very large PGR efficiencies---up to several \unit{\tera\hertz\per\milli\watt} per resonant mode---can be achieved at triple resonance, several orders of magnitude larger than current state-of-the-art sources. However, because achieving triple resonance in practice is difficult, we consider the off-resonance PGR properties and optimize for the maximum detuning case. It is clear that higher pump coupling increases the maximum detuning PGR, so our resonator's pulley coupler should have a high coupling ratio for the pump.

Following the above discussion, we simulated a racetrack resonator source and calculated its figures of merit. We chose to use 1 mm long straight waveguide sections for the racetrack and we periodically pole one side. The length is chosen to increase PGR efficiency while not being too long such that manufacturing variance on a wafer would significantly affect the waveguide shape. The resonator simulations calculate an average bandwidth of \SI{0.513}{\giga\hertz} and average FSR of \SI{46.9}{\giga\hertz}, which is a large enough difference for the produced islands to have minimal overlap and produce high-fidelity entanglement. We choose the pump laser to have a bandwidth of \SI{5}{\giga\hertz}, and we assume a Fourier transform-limited Gaussian spectrum such that it has pulse time width of 88.2 ps, pulse energy of \SI{1.11}{\pico\joule}, and a repetition rate of \SI{900}{\mega\hertz}. This choice of parameters represents a setup with 1~mW average pump power and an estimated per-island pair-generation probability per pulse of approximately $1.84 \times 10^{-2}$; under the ZALM protocol this corresponds to a heralded perfectly entangled frequency-state fidelity above 0.998 after accounting for the factor-of-two increase from deterministic wavelength-based splitting~\cite{ZALM}.

Our simulations predict 106 well-defined spectral islands over the 190.95--\SI{195.94}{\tera\hertz} interval, with a total internal PGR efficiency of \SI{1.75}{\giga\hertz\per\milli\watt}. Of these pairs, a fraction of 0.616 is coupled out of the cavity with both photons, giving an external PGR efficiency of \SI{1.08}{\giga\hertz\per\milli\watt}. The mean per-island PGR is \SI{16.55}{\mega\hertz\per\milli\watt} with a standard deviation of \SI{0.042}{\mega\hertz\per\milli\watt}, spanning \SI{16.49}{\mega\hertz\per\milli\watt} to \SI{16.63}{\mega\hertz\per\milli\watt}. The corresponding average brightness is $1.85 \times 10^1$~\unit{\mega\hertz\per\milli\watt\per\giga\hertz}. The Schmidt decomposition evaluated over the full JSA frequency interval gives a von Neumann entropy of 4.66 and an effective frequency-space Schmidt number of 105.99 for the 106-island state. Applying the same numerical resolution to each island separately gives an average single-island purity of 0.994 with a standard deviation of $5.7 \times 10^{-5}$. Following deterministic wavelength-based splitting, we fold the spectrum about the half-pump frequency and recalculate the Schmidt decomposition for the resulting 53 logical frequency-bin pairs, obtaining an effective frequency-space Schmidt number of 53.00. The slight departure from exactly one-half of the unsplit value results from the modified distribution of island weights after folding and normalization.

\section{Conclusions}

We have presented a simulation-based design study of an integrated photonic racetrack resonator SPDC source for frequency-multiplexed, frequency-entangled photon pair generation. The central result is a simulated set of 106 doubly resonant signal and idler frequency-mode pairs with an effective frequency-space Schmidt number of 105.99, an average bandwidth of \SI{0.513}{\giga\hertz}, an FSR of \SI{46.9}{\giga\hertz}, and a total internal PGR efficiency of \SI{1.75}{\giga\hertz\per\milli\watt}. Under deterministic wavelength splitting, the folded 53-pair state has an effective frequency-space Schmidt number of 53.00.

The principal theoretical contribution is a closed-form derivation that maps classical resonator parameters directly to the quantum joint spectral amplitude and pair generation rate, using the Raymer dispersive-medium quantization formalism as its Hamiltonian foundation. Because this starting point accounts for waveguide dispersion and proper field normalization---unlike the plane-wave treatments used in prior cavity JSA derivations~\cite{jsa-shape-theory-Jeronimo-Moreno2010}---it yields formulas that are predictive for cavity-derived quantum figures of merit within the accuracy of the underlying electromagnetic model (see discussion of segmented EME phase error in Sec.~\ref{sec:simulation} and Appendix~\ref{sec:AppD-segmented-EME}). This enables the classical electromagnetic simulation pipeline presented here to serve as a pre-fabrication design tool that predicts quantum figures of merit, not just classical resonator properties.

The racetrack geometry with straight-section-only periodic poling, adapted from prior SHG and OPO work~\cite{mckenna2022ultralow, park2024squeezed, hwang2024spdc}, enables compatibility with x-cut TFLN substrates while allowing straightforward adjustment of the FSR, phase-matching bandwidth, and PGR efficiency through changes to the straight section length. Key design trade-offs are the phase-matching bandwidth vs PGR efficiency (both set by the poled section length), the cavity Q vs out-coupling efficiency (set by the directional coupler gap), and the number of supported islands vs per-island efficiency.

The primary limitation of the current design is operation with a nonresonant pump, which is a practical consequence of the double-resonance condition. Achieving triple resonance through active tuning would increase the PGR by several orders of magnitude and represents the most impactful avenue for future work. Additional future directions include experimental fabrication and characterization of the device, extension of the simulation pipeline to include coupling losses and detector efficiencies for end-to-end network rate predictions, and investigation of the design on alternative $\chi^{(2)}$ platforms.

\begin{acknowledgments}
The authors acknowledge useful discussions with Dr. Brian Smith of University of Oregon, and with Siavash Mirzaei Ghormish of Brigham Young University. The authors gratefully acknowledge funding from the National Science Foundation Engineering Research Center for Quantum Networks under Grant No. 1941583.
\end{acknowledgments}

\section*{Data Availability}
The data that support the findings of this article are not publicly available but may be obtained from the authors upon reasonable request.

\appendix

\section{Derivation of the SPDC Joint Spectral Amplitude Formula for a Straight Waveguide}\label{sec:AppA-straightwgJSAderiv}

Here we present an abridged version of a recently developed theory of multi-longitudinal-mode cavity-enhanced SPDC, the purpose of which is to support our estimates of the achievable rate of photon pair generation in a series of spectrally separable phase-matched regions (``islands''). The distinguishing features of this treatment relative to prior cavity JSA derivations~\cite{jsa-shape-theory-Jeronimo-Moreno2010} are: (1)~the use of the Raymer dispersive-medium quantization formalism~\cite{Raymer20-dispersive-quantization} as the Hamiltonian foundation, which properly accounts for waveguide dispersion and field normalization in a spatially structured dielectric medium; and (2)~the derivation of the joint spectral amplitude spanning all spectral islands in the presence of cavity boundary conditions. The Hamiltonian starting point is taken from Raymer's framework rather than derived from first principles here; that derivation will be presented in a follow-up paper. All formulas needed to implement the simulation pipeline are self-contained in this appendix.

The starting point of such a theory has to be a correct formulation of a Hamiltonian for the energy stored in a spatially structured medium having spectral dispersion, for which we use the formalism developed by Raymer~\cite{Raymer20-dispersive-quantization, Quesada-22-beyond-photon-pairs, Drummond-Hillery-2014}. The first-principles quantum mechanical Hamiltonian, when adapted to a dielectric waveguide structure, consists of two parts---a linear energy for the SPDC field and a nonlinear energy describing a type-0 second-order nonlinear interaction in which all fields are co-polarized and quasi-phase matching is achieved by spatial modulation of the nonlinearity,
\begin{equation}\label{eq:H1}
    H^{(1)} = \sum_{j} \int_{0}^{\infty} \frac{d\omega}{2\pi} \hbar\omega \hat{b}_j^\dagger(\omega) \hat{b}_j(\omega)\text{,}
\end{equation}
where the first-order Hamiltonian is summed over all transverse modes for the signal and idler photons $j$, and the second-order Hamiltonian
\begin{equation}\label{eq:H2}\begin{split}
H^{(2)} = \sum_{j,k} & \int_{-\infty}^{\infty} \int_{-\infty}^{\infty}\int_{-\infty}^{\infty} \frac{d\omega}{2\pi} \frac{d\omega'}{2\pi} \frac{d\omega_p}{2\pi}\\
 &e^{-i\Delta \omega t} 
 G_{jk}(\omega, \omega', \omega_p)\ \hat{b}_j^\dagger(\omega) \hat{b}_j^\dagger(\omega') \hat{b}_k(\omega_p) + \text{H.c.}
\end{split}\end{equation}
is summed over both the transverse waveguide modes of the telecom frequencies $j$ and the modes of the pump frequencies $k$. We assume a nondepleting pump that can be treated as a constant classical field so we omit its first-order contribution to the Hamiltonian. We are only considering degenerate type-0 SPDC where the signal and idler photons are in the same transverse waveguide mode and have frequencies $\omega$ and $\omega'$. In the above formula, $\Delta\omega = \omega + \omega' - \omega_p$ and $\hat{b}_m(\omega_n)$ is the annihilation operator for polaritons (mixed photonic and material excitations in a nonresonant medium) in a transverse waveguide mode labeled by $m$ with wave number $\beta_m(\omega_n)$ and energy $\hbar\omega_n$. We assume all modes $G_{jk}$ are TE-polarized. The pump mode is labeled $k$ and the signal and idler modes are labeled $j$.

The commutation relation between the ladder operators is
\begin{equation}
    \left[\hat{b}_m(\omega), \hat{b}_n^\dagger(\omega') \right] = 2\pi\delta(\omega - \omega')\delta_{nm}\text{.}
\end{equation}
Using numerical solvers, we find the transverse waveguide modes $W_n(\mathbf{x}, \beta_n(\omega))$ which are orthogonal for equal angular frequencies under the integral
\begin{equation}
\int d^2x \rho_J(\mathbf{x}) W_n^*(\mathbf{x},\beta_n(\omega))  W_m(\mathbf{x},\beta_m(\omega)) = \delta_{nm}\text{,}
\end{equation}
with the weight function approximated as a function only of position within a given frequency band (i.e., visible and telecom),
\begin{equation}\label{eq:weight-func}
\rho_J(x) = \varepsilon_0 \left\{ \eta(r,\omega) - \frac{\omega}{2} \frac{\partial\eta(r,\omega)}{\partial\omega} \right\}_{\omega=\omega_J}\text{,}
\end{equation}
where $\eta(r,\omega)$ is the inverse of the linear susceptibility $\eta(r,\omega) = \left[\varepsilon(r,\omega)\right]^{-1}$, and $\omega_J$ is the center of the frequency band being considered. This “band approximation” is commonly used to model waveguide modes, but misses some typically small corrections that arise when considering that
the exact modes of a dispersive waveguide structure are not strictly orthogonal~\cite{Raymer20-dispersive-quantization}. These corrections will be considered in future studies. We also note that Lumerical's mode solver uses Poynting vector orthogonality, which does not include the second term in Eq.~(\ref{eq:weight-func}). However, for weakly dispersive waveguides omitting the second term is a reasonable approximation.

 We consider only type-0 SPDC where all optical fields are co-polarized in TE modes, so we can take the fields to be approximated as scalar functions. Focusing on a single transverse mode for the generated SPDC, the field that is properly quantized in dielectric media is the electric displacement field, which represents the combination of light and material excitations. We can write the positive-frequency part of the displacement field operator in a given band $B_J$ as
\begin{equation}
\hat{D}_j^{(+)}(\mathbf{r},t) = i \int_{B_j} \frac{d\omega}{2\pi} \theta_j(\omega) \hat{b}_j(\omega) e^{i\left[\beta_j(\omega) z - \omega t\right]} W(\mathbf{x},\beta_j(\omega))\text{,}
\end{equation}
where $\theta_j(\omega) = \sqrt{\frac{\varepsilon_0\ \hbar\ \omega}{2\ v_{g,j}(\omega)}}$, where $v_{g,j}$ is the group velocity for waveguide mode $j$. Quantization of the electric field instead of the displacement field can lead to errors and is fundamentally inconsistent with quantum mechanics when treating the system in terms of the macroscopic (effective-field) Maxwell equations~\cite{Drummond-Hillery-2014, Quesada-22-beyond-photon-pairs}. Our use of frequency-based operators rather than position-localized ones (as is often used in such derivations) is required by the need to quantize field operators extending throughout the entire volume of the medium to properly handle dispersive media, as there is no spatially and spectrally localized field operator that can handle dispersion when treating the medium as transparent~\cite{Loudon-qoptics, Milloni-qoptics, Drummond-Hillery-2014, Raymer20-dispersive-quantization}.

The simple forms for the Hamiltonian in Eqs.~(\ref{eq:H1}) and (\ref{eq:H2}) are the result of in-depth considerations that ensure consistency with the classical Maxwell equations for a transparent medium and quantum mechanics. Quantum nonlinear optics theory yields an expression for the nonlinear coupling coefficient,
\begin{widetext}\begin{equation}
G_{jk}(\omega,\omega',\omega_p) = i\ C\ \frac{2}{\pi} \int_V d^2\mathbf{x}\ dz \, F^{(2)}\left(\mathbf{x},\omega_p;\omega,\omega' \right) W_j^*(\mathbf{x},\beta_j(\omega)) W_j^*(\mathbf{x},\beta_j(\omega')) W_k(\mathbf{x},\beta_k(\omega_p)) e^{i \Delta \beta z}\text{,}
\end{equation}\end{widetext}
where 
\begin{align*}
    \Delta \beta &= \beta_k(\omega_p)  - \beta_j(\omega) - \beta_j(\omega') - \frac{\pi}{\Lambda}\text{,}\\
    C &= 2\sqrt{ \left(\varepsilon_0\frac{\hbar}{2}\right)^3 \frac{\omega\ \omega'\ \omega_p}{v_{g,j}(\omega) v_{g,j}(\omega') v_{g,k}(\omega_p)}}\text{,} \\
    F(\mathbf{x},\omega_p;\omega,\omega') &\approx \eta^{(2)}(\mathbf{x},\omega_p;\omega,\omega')\\
    &= \frac{\varepsilon_0 \chi^{(2)}\left(\mathbf{x},\omega_p;\omega,\omega'\right)}{\left[\varepsilon^{(1)}(\omega) \varepsilon^{(1)}(\omega') \varepsilon^{(1)}(\omega_p)  \right]}\text{,}\\
\end{align*}
where the factor of $2/\pi$ is introduced when we substitute the square-wave modulation of the nonlinearity (using quasi-phase matching) with an exponential factor $e^{i \pi z/\Lambda}$ in the phase mismatch. The constants each mean the following: $\Delta \beta$ is the phase mismatch between the various SPDC modes and $\Lambda$ is the poling period of the waveguide, $C$ collects the normalization constants from the displacement field normalization factors, and $F$ is the nonlinear interaction coefficient, which can be approximated as the second-order inverse permittivity $\eta^{(2)}$. We note that a fully rigorous treatment in a dispersive, structured medium would include extra terms in the formula for $F$ but these extra terms are small and can be neglected for our purposes. A full derivation of the exact nonlinear coefficient may be introduced in future work.

We can now make a few assumptions to simplify the above expression. First, we assume that the nonlinear susceptibility can be replaced with a geometry-independent effective susceptibility for the waveguide geometry under consideration $\chi^{(2)}(\mathbf{x},\omega_p;\omega,\omega') \rightarrow \chi^{(2)}_{\text{eff}}(\omega_p;\omega,\omega')$, which only depends on the waveguide material. Second, we assume that each interacting SPDC mode is in one specific waveguide mode (in our case they are all in the fundamental TE mode) so we can drop the waveguide mode index: $\beta_j(\omega) \rightarrow \beta(\omega)$, $v_{g,j}(\omega) \rightarrow v_g(\omega)$. We note that this approximation is not applicable to all SPDC processes in general. In our case, the type-0 process, the fact that all three SPDC modes are assumed to be in the same spatial mode, and the opposite-island JSA means both signal and idler photons are in the exact same state and even their marginal frequency distributions are the same. The only difference is noticed when the joint frequency state is measured. Therefore, the frequency variable $\omega, \omega'$ is sufficient to indicate the ``signal and idler'' photons uniquely (though technically there is no distinguishing feature of the photon designated with $\omega$ vs $\omega'$ unless they are designated using some artificial metric such as path after separation or saying all photons below the half pump frequency are signal and photons with higher frequency are idler). Third, we expand the first-order permittivities to use the effective refractive indices of the corresponding modes $\varepsilon^{(1)}(\omega) = \varepsilon_0n^2(\omega)$. With these simplifications, we can rewrite the nonlinear coupling coefficient as
\begin{equation}\begin{split}
    G(\omega,\omega', \omega_p) & = i \frac{\sqrt{2}\ \hbar^{\frac{3}{2}}L}{\pi\sqrt{\varepsilon_0}}   \sqrt{\frac{\omega\ \omega'\ (\omega_p)}{v_g(\omega) v_g(\omega') v_{g}(\omega_p) }}\\
      \times\ & O(\omega,\omega', \omega_p)\ \frac{\chi^{(2)}\left(\omega_p; \omega, \omega' \right)}{n^2(\omega) n^2(\omega') n^2(\omega_p)} \\
      &\times  \text{sinc}\left(\frac{\Delta \beta\ L}{2}\right)\text{,}
\end{split}\end{equation}
where we have defined the mode overlap integral function
\begin{equation}\begin{split}
O(\omega,\omega', \omega_p) \equiv \int_A d^2\mathbf{x}\ W^* &\left( \mathbf{x},\beta(\omega) \right) W^*\left( \mathbf{x},\beta(\omega') \right)\\
\times & W\left(\mathbf{x},\beta(\omega_p)\right)\text{.}
\end{split}\end{equation}

We can now write the evolution of the ladder operators in a straight waveguide SPDC process. An SPDC process driven by a pulsed pump field can be treated as a scattering problem in which only input and output states are considered, not the detailed space-time evolution of fields and state during the scattering process. Using the above Hamiltonians, the Heisenberg-picture time evolution is given by
\begin{equation}\begin{split}
    \partial_t \hat{b}(\omega) &= \frac{i}{\hbar}[H^{(2)}, \hat{b}(\omega)]\\
    &\begin{split}
    =  \frac{i}{\hbar} \int_{-\infty}^{\infty} \int_{-\infty}^{\infty} &\frac{d\omega_p}{2\pi}   \frac{d\omega'}{2\pi} G(\omega, \omega', \omega_p)\\
    & e^{-i\Delta \omega t} \alpha_p(\omega_p)\ \hat{b}^\dagger(\omega')\text{,}
    \end{split}
\end{split}\end{equation}
where we replaced the operator $\hat{a}_p$ with $\alpha_p$, the coherent-state (``classical'') amplitude of the pump. We integrate the equation over the travel time $T$ for the pump pulse to travel through the periodically poled waveguide, evolving from time $-T/2$ to $+T/2$, which gives us the following formula for the output mode operator
\begin{equation}\begin{split}
    \hat{b}_{\text{out}}(\omega) \approx \hat{b}_{\text{in}}(\omega)
    &+  \\
    \frac{i}{\hbar} \int_{-T/2}^{T/2} &dt \, \int_{-\infty}^{\infty} \int_{-\infty}^{\infty} \frac{d\omega_p}{2\pi}  \frac{d\omega'}{2\pi} \\
    &G (\omega,\omega', \omega_p)\ e^{-i\Delta \omega t}\  \alpha_p(\omega_p)\ \hat{b}_{\text{in}}^\dagger(\omega')\text{,}
\end{split}\end{equation}
where we have implicitly carried out the integration over $\omega$ using the delta function from the commutator. We can then carry out the time integration to yield an approximate delta function, enforcing energy conservation such that $\omega_p = \omega + \omega'$. This gets us the output formula
\begin{equation}\begin{split}
    \hat{b}_{\text{out}}(\omega) \approx \hat{b}_{\text{in}}(\omega) + &\\
    \int_{0}^{\infty} \frac{d\omega'}{2\pi} &J(\omega,\omega') \hat{b}_{\text{in}}^\dagger(\omega')\text{,}
\end{split}\end{equation}
where
\begin{equation}
J(\omega,\omega') = \frac{i}{\hbar} G(\omega,\omega', \omega+\omega')\ \alpha_p(\omega + \omega')\text{.}
\end{equation}
The function $J(\omega,\omega')$ can be thought of as a Green's function or a propagator for the nonlinear interaction through the waveguide. One can show that $J(\omega,\omega')$ is essentially the joint-spectral amplitude describing the two-photon state generated by  SPDC without a cavity~\cite{Raymer05-pure-state-generation, Quesada-22-beyond-photon-pairs}.

\section{Derivation of the Cavity-enhanced JSA and PGR}\label{sec:AppB-cavityJSAderiv}

Armed with the ladder operators for a straight waveguide SPDC source, we now derive the output operator for a cavity source. We define the cavity field ladder operators to be $\hat{c}\text{,}\ \hat{c}^\dagger$ with the commutation relation $[\hat{c}_m(\omega), \hat{c}_n^\dagger(\omega')] = 2\pi\delta(\omega-\omega')\delta_{nm}$. When the waveguide field arrives on the cavity interface, it exchanges energy with the displacement field outside the cavity, whose operator form is
\begin{equation}\begin{split}
    \hat{D}_{\text{out},j}^{(+)}(\mathbf{r},t) = i &\int_{B_j}  \frac{d\omega}{2\pi} \sqrt{\frac{\varepsilon_0 \hbar \omega}{2\ v_{g,j}(\omega)}} \hat{a}_j(\omega)\\
    &\times W_{\text{out},j}(\mathbf{x},\beta_j(\omega)) e^{i(\beta_j(\omega)z - \omega t)}\text{.}
\end{split}\end{equation}
Boundary conditions at the coupler are expressed by the standard form, valid in classical and quantum theory,
\begin{align}
\hat{c}(0^+,\omega) &= \sigma \hat{c}(L^-,\omega) + i\kappa \hat{a}_{in}(\omega)\text{,} \\
\hat{a}_{out}(\omega) &= i\kappa \hat{c}(L^-,\omega) + \sigma \hat{a}_{in}(\omega)\text{,}
\end{align}
where $\sigma$ and $\kappa = (1-\sigma^2)^{1/2}$ are the real amplitude reflectivity and transmissivity of the coupler, and $\hat{a}_{in}(\omega), \hat{c}(0^+,\omega), \hat{c}(L^-,\omega), \hat{a}_{out}(\omega)$ are, respectively, the annihilation operators for the fields just before the input coupler on the input side, inside just after the input coupler, inside after a cavity round trip, and outside just after the coupler~\cite{Raymer13-quantum-cavity-IO}. Thus $\hat{a}_{in}(\omega), \hat{a}_{out}(\omega)$ are the input and output operators for the cavity as a whole. These boundary conditions represent the conservation of polariton number across a lossless interface. We can combine the boundary conditions to get the output field operator as a function of the cavity field and input field,
\begin{equation}
    \hat{a}_{out}(\omega) = \frac{1}{\sigma}\hat{a}_{in}(\omega) + i \frac{\kappa}{\sigma} \hat{c}(0^{+},\omega)\text{.}
\end{equation}

Including the effects of the nonlinear medium and a phenomenological damping term $\eta$, the field operator evolves upon a single round trip as
\begin{equation}\begin{split}
    \hat{c}(L^-,\omega) = \eta  & e^{i\beta(\omega)L} \left[ \hat{c}(0^+,\omega) \right.\\
    & +\left. \int_0^\infty \frac{d\omega'}{2\pi} \, J(\omega,\omega') \hat{c}^\dagger(0^+,\omega') \right]\text{.}
\end{split}\end{equation}
Combining with the boundary conditions yields
\begin{equation}\label{eq:recursive-evol}\begin{split}
    \hat{c}(0^+,\omega&) = \frac{i\kappa}{1-\sigma\eta e^{i\beta(\omega) L}}  \hat{a}_{in}(\omega)\\
     &+ \int_0^\infty \frac{d\omega'}{2\pi} \, \frac{\sigma\eta e^{i\beta(\omega) L} }{1-\sigma\eta e^{i\beta(\omega) L}}  J(\omega,\omega')\hat{c}^\dagger(0^+,\omega') \text{.}
\end{split}\end{equation}
The lowest-order perturbative solution, valid for SPDC, is obtained by making the substitution
\begin{equation}
\hat{c}^\dagger(0^+,\omega) \to \frac{-i\kappa}{1-\sigma\eta e^{-i\beta(\omega) L}} \hat{a}_{in}^\dagger(\omega)\text{,}
\end{equation}
inside the integral. Inserting into Eq.~(\ref{eq:recursive-evol}) gives the sought solution,
\begin{equation}\label{eq:linearized-evol}\begin{split}
    \hat{a}_{out}(\omega) = h(\omega)&\hat{a}_{in}(\omega) +\\
    &e^{i\beta(\omega)L} \int_0^\infty \frac{d\omega'}{2\pi} \, j(\omega,\omega') \hat{a}_{in}^\dagger(\omega')\text{,}
\end{split}\end{equation}
where the functions $h$, $j$ are given by
\begin{align}
h(\omega)  &= -e^{i\beta(\omega)L} \frac{\eta - \sigma e^{-i\beta(\omega)L}}{1-\sigma\eta e^{i\beta(\omega)L}}\text{,}\\
j(\omega,\omega') &= \frac{\kappa^2 \eta J(\omega,\omega')}{(1-\sigma\eta e^{i\beta(\omega)L})(1-\sigma\eta e^{-i\beta(\omega')L})}\text{.}
\end{align}

Thus, the spectral density function equals the waveguide source's spectral density multiplied by the cavity density of states for both emitted photons. A result similar in form to this one has been derived in Jeronimo-Moreno \textit{et al.}, but using plane waves rather than modes of a dispersive dielectric waveguide as needed for a quantitative calculation of the expected photon-pair generation rate in such structures~\cite{jsa-shape-theory-Jeronimo-Moreno2010}. The factor $h(\omega)$ in the first term of Eq.~(\ref{eq:linearized-evol}) gives the complex phase factor accumulated by any input pulse incident on the cavity, and in the case of an empty cavity without a loss factor reproduces the well-known result for phase accumulation in a lossless cavity~\cite{Raymer13-quantum-cavity-IO}.
 
The final step is to find the pair generation probability of the produced photon pairs from the operator solution above. The differential probability of generating a photon pair from vacuum input in degenerate SPDC can be calculated by dividing the average photon number count at frequency $\omega$ by 2 to account for the fact that the signal and idler photons have complementary frequency spectra. We note that this formula assumes that we are operating in the low photon number regime and is not generally valid when multiphoton pair production events occur with relevant frequency.
\begin{equation}
    dP_{pair}(\omega) \approx \frac{1}{2} \braket{\hat{n}(\omega)} = \frac{1}{2} \braket{0_{in}| \hat{a}^\dagger_{out}(\omega) \hat{a}_{out}(\omega) |0_{in}}\text{.}
\end{equation}
We take the expectation values above between the vacuum input state. We can then expand the output operators as functions of the input field operators, noting that only the $a_{in}a^\dagger_{in}$ terms evaluate to nonzero results.
\begin{equation}\begin{split}
    \braket{\hat{n}(\omega)} = \bra{0}  &\int_0^\infty \int_0^\infty \frac{d\omega'}{2\pi} \frac{d\omega''}{2\pi}\\ 
     \times &j^*(\omega,\omega') j(\omega,\omega'') \hat{a}_{in}(\omega')\hat{a}_{in}^\dagger(\omega'') \ket{0}\text{.}
\end{split}\end{equation}
Next we use the identity $\left[\hat{a}(\omega),\hat{a}^\dagger(\omega') \right] = 2\pi \delta(\omega-\omega')$, which lets us write
\begin{equation}
    \hat{a}(\omega) \hat{a}^\dagger(\omega') = \hat{a}^\dagger(\omega')\hat{a}(\omega) + 2\pi \delta(\omega-\omega')\text{.}
\end{equation}
Then, since $\hat{a}^\dagger(\omega') \hat{a}(\omega) \ket{0} = 0$, $\hat{a}(\omega) \hat{a}^\dagger(\omega') \ket{0} = 2\pi \delta(\omega-\omega')$. This lets us evaluate the inner product to
\begin{align}\label{eq:avg-n}
    \braket{\hat{n}(\omega)} =  \int_0^\infty &\int_0^\infty \frac{d\omega'}{2\pi} \frac{d\omega''}{2\pi} j^*(\omega,\omega')\nonumber\\
    & \times j(\omega,\omega'') 2\pi \delta(\omega'-\omega'')\\
    = \int_0^\infty & \frac{d\omega'}{2\pi} |j(\omega,\omega')|^2\nonumber\text{.}
\end{align}
We note that this formula is simply the marginal probability density distribution for one of the photons in the pair with frequency $\omega$. We can use this formula to infer that the spectral probability density for this state is
\begin{equation}
    |\Psi(\omega,\omega')|^2 = \frac{1}{2} \left|j(\omega,\omega') \right|^2\text{.}
\end{equation}

We are interested in calculating the probability of observing a photon pair in a given joint spectral interval of ``area'' $\Omega^2$. For any single pulse this probability is given by
\begin{equation}\label{eq:PGR-cavity}
P_{\text{total}} = \int_{-\Omega/2}^{\Omega/2} \int_{-\Omega/2}^{\Omega/2} \frac{d\omega}{2\pi} \frac{d\omega'}{2\pi} \left|\Psi(\omega,\omega')\right|^2\text{.}
\end{equation}
If we take $\Omega \rightarrow \Omega_{FSR}$ to be the free spectral range of the cavity, then $P$ is the probability to observe a photon pair in one of the joint spectral islands.
\begin{equation}
P_{\text{island}} = \int_{-\Omega_{FSR}/2}^{\Omega_{FSR}/2}  \int_{-\Omega_{FSR}/2}^{\Omega_{FSR}/2} \frac{d\omega}{2\pi} \frac{d\omega'}{2\pi} \left|\Psi(\omega,\omega')\right|^2\text{.}
\end{equation}
The same probability in the absence of a cavity ($\kappa = 1, \sigma = 0$) is
\begin{equation}\label{eq:PGR-waveguide}
P_{\text{no cavity}} = \int_{-\Omega_{FSR}/2}^{\Omega_{FSR}/2} \int_{-\Omega_{FSR}/2}^{\Omega_{FSR}/2} \frac{d\omega}{2\pi} \frac{d\omega'}{2\pi} \left| \eta^2 J(\omega,\omega') \right|^2\text{,}
\end{equation}
which gives a clear way to define and calculate the probability of photon pair generation per pump pulse into any single spectral island with or without a cavity.

We can then calculate the expected pair generation rate from a cavity source by making assumptions about the pump pulse frequency spectrum, pulse energy, and pump repetition rate $r$,
\begin{equation}
    R = r P_{\text{total}}\text{,}
\end{equation}
where we set the pump coherent-state amplitude using $\alpha_p (\omega + \omega') =  \sqrt{\braket{\hat{n}_p}} = \sqrt{ \int_{-T/2}^{T/2} dt  P_p(t) \Gamma(\omega + \omega') e^{i} / \hbar\ \omega_p }$ where $P(t)$ is the power of the pump pulse at time t and $\Gamma(\omega_p)$ is the frequency spectrum of the pump pulse normalized such that $\int d \omega_p\ \Gamma(\omega_p) = 1$. We note that as long as we are in the low time-averaged photon number regime and because the pair generation rate depends on $|\alpha_p|^2 \sim P_p$, the average number of pairs generated over all time does not depend on the instantaneous pump power, only the time-averaged pump power. Therefore, it is straightforward to compare our pulsed cavity source to sources in the literature---which commonly assume continuous-wave pump lasers---by setting the product of the pump repetition rate and the pump pulse energy $r \mathcal{E}_p$ equal to 1 mW.

Finally, since many cavity source papers list the in-cavity (internal) PGR efficiency as a way to characterize the nonlinear conversion efficiency of the source, we note that we can calculate the in-cavity PGR for our source by making the substitution $j(\omega,\omega') \rightarrow j'(\omega,\omega') = j(\omega,\omega')/\kappa$ in Eq.~(\ref{eq:avg-n}) to remove the effect of coupling out of the cavity.

\section{Cavity Brightness Enhancement}\label{sec:AppC-cavity-enhancement}

From Eqs.~(\ref{eq:PGR-cavity}) and (\ref{eq:PGR-waveguide}), we can calculate the enhancement of photon pair generation by the cavity. We separate the round-trip loss into the losses due to the nonlinear section $\eta_{NL}$ and the loss due to the rest of the cavity $\eta_{cav}$, where $\eta = \eta_{NL}\ \eta_{cav}$, and then expand the island probability amplitude as 
\begin{equation}
    \left| \Psi(\omega,\omega') \right|^2 = \frac{\kappa^4\ \eta_{NL}^2\ \eta_{cav}^2}{2} f(\omega) f^*(\omega')\left| J(\omega,\omega')\right|^2\text{,}
\end{equation}
where $f(\omega)$ is the cavity filter function
\begin{equation}
    f(\omega) = \frac{1}{\left|1-\sigma\ \eta_{NL}\ \eta_{cav} e^{i\beta(\omega)L}\right|^2}\text{.}
\end{equation}
Next, we rewrite the cavity filter function in the more familiar form
\begin{equation}
    f(\omega) = \frac{1}{(1-\eta_{NL}\ \eta_{cav} \sigma)^2} \frac{1}{1 + \left(\frac{2 F}{\pi} \right)^2 \text{sin}^2\left( \frac{\beta(\omega) L}{2}\right)}\text{,}
\end{equation}
where $F$ is the cavity finesse, given by the familiar form $F = \frac{\pi \sqrt{\eta_{NL}\ \eta_{cav}\ \sigma}}{1-\eta_{NL}\ \eta_{cav}\ \sigma}$. When loss is neglected, this result is equivalent to that of Jeronimo-Moreno \textit{et al.} derived using different methods~\cite{jsa-shape-theory-Jeronimo-Moreno2010}.

The peak value of the joint spectral intensity occurs on a signal-idler double resonance when $e^{i \beta(\omega) L} = 1$. On a double resonance, the maximum value is
\begin{equation}
|\Psi|_{max}^2 = \frac{\kappa^4\ \eta_{NL}^2\ \eta_{cav}^2}{\left(1-\eta_{NL}\ \eta_{cav}\ \sigma \right)^4} \frac{\left| J_{max} \right|^2}{2}\text{,}
\end{equation}
and the full width at half maximum (FWHM) of a cavity resonance for one of the frequency variables is
\begin{equation}\begin{split}
    \Gamma =& \frac{4}{T} \text{arcsin}\left( \frac{1 - \eta_{NL}\ \eta_{cav}\ \sigma}{2\sqrt{\eta_{NL}\ \eta_{cav}\ \sigma}} \right)\\
    =& \frac{4}{T}\text{arcsin}\left(\frac{\pi}{2 F} \right) \approx \frac{1}{T} \frac{2 \pi}{F}\text{,}
\end{split}\end{equation}
where the final approximation is valid for large cavity finesse. We can then calculate the probability for photon pair generation within a single FSR of the cavity $\{-\pi/T, \pi/T\}$ where T is the cavity round trip time for the specific resonance.
\begin{equation}\begin{split}
    P_{cav} =& \int_{-\pi/T}^{\pi/T} \int_{-\pi/T}^{\pi/T} \frac{d\omega}{2\pi} \frac{d\omega'}{2\pi} |\Psi(\omega,\omega')|^2\\
    =& \left(\frac{\kappa^2\ \eta_{NL}\ \eta_{cav}}{T(1- \sigma^2\ \eta_{NL}^2\ \eta_{cav}^2)} \right)^2 \frac{|J_{max}|^2}{2}\text{.}
\end{split}\end{equation}

\begin{figure}[!htbp]
    \centering
    \includegraphics[width=0.95\linewidth]{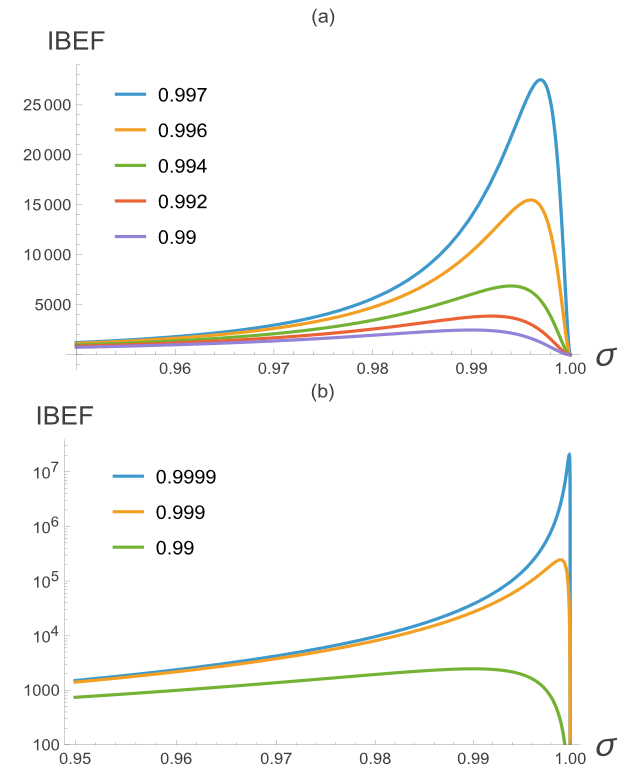}
    \caption{Integrated brightness enhancement factor (IBEF) as a function of single-pass reflection coefficient $\sigma$ for various loss values. Legends show the value of $\eta_{cav}$ while $\eta_{NL}=0.99999$ is fixed. Plot (a) shows that the maximum IBEF point happens at higher reflectivity for lower losses, while plot (b) illustrates that each order of magnitude decrease in round trip losses corresponds to roughly 2 orders of magnitude increase in enhancement factor.}\label{fig:brightness-enhancement}
\end{figure}

To understand how strongly the cavity enhances SPDC, first note that in the limit of no cavity ($\sigma=0, \eta_{cav} = 1$) the cavity JSI reduces to the free-space JSI, with nonlinear-medium loss accounted for,
\begin{equation}
    |\Psi(\omega,\omega')|_{wg}^2 = \eta_{NL}^2 \frac{|J_{max}|^2}{2} \text{.}
\end{equation}
To create a photon pair state from a straight waveguide source that matches the cavity state as closely as possible, one would pass the broadband SPDC from a noncavity source through a spectral filter with a transmission function that mimics a cavity resonance line shape. Consider the light within a single FSR transmitted through such a filter. The JSI of that filtered light will be given by
\begin{equation}\begin{split}
    |\Psi (\omega,\omega')|_{fwg}^2 = |\Psi (\omega,\omega')|^2_{wg} &\\
    \times f(\omega) & f^*(\omega') H_n(\omega)  H_m(\omega')\text{,}
\end{split}\end{equation}
where $H_n$ is a top-hat function in the FSR interval around the resonance $n$
\begin{equation}
 H_n(\omega) =  \begin{cases} 
              1 & n/T\  \leq \frac{\beta(\omega) L}{2\pi} < (n + 1)/T , n \in \mathbb{Z} \\
              0 & \text{otherwise}
           \end{cases}\text{.}
\end{equation}

Next, we evaluate the probability integral for the filtered straight waveguide source case.
\begin{equation}\begin{split}
    P_{fwg} = \int \int_{-\pi/T}^{\pi/T} \frac{d\omega}{2\pi} & \frac{d\omega'}{2\pi}\\
    \times \eta_{NL}^2 &\frac{\left| J(\omega, \omega') \right|^2}{2} f(\omega) f^*(\omega')\text{.}
\end{split}\end{equation}
We can approximate the above integral as
\begin{equation}
    P_{fwg} \approx \eta_{NL}^2 \frac{(1-\eta_{NL}\ \eta_{cav}\ \sigma)^2}{T^2(1 + \eta_{NL}\ \eta_{cav}\ \sigma)^2} \frac{\left|J_{max} \right|^2}{2}\text{.}
\end{equation}

Finally, we can evaluate the IBEF between the two source configurations.
\begin{equation}
    \text{IBEF} = \frac{P_{cav}}{P_{fwg}} = \frac{\eta_{cav}^2 \kappa^4}{(1-\eta_{NL}\ \eta_{cav}\ \sigma)^4}\text{.}
\end{equation}
As illustrated in Fig.~\ref{fig:brightness-enhancement}, the cavity enhancement reaches its maximum value when $\sigma = \eta$, i.e., when the produced photon pairs are critically coupled. The maximum value of IBEF is then
\begin{equation}
    \text{IBEF}_{max} = \frac{\eta_{cav}^2}{(1-\eta_{cav}^2\ \eta_{NL}^2)^2}\text{.}
\end{equation}

\section{Approximation of Propagation Loss in Large Bends Using an Eigenmode Expansion Method}\label{sec:AppD-segmented-EME}

\begin{figure}[!t]
    \centering
    \includegraphics[width=1\linewidth]{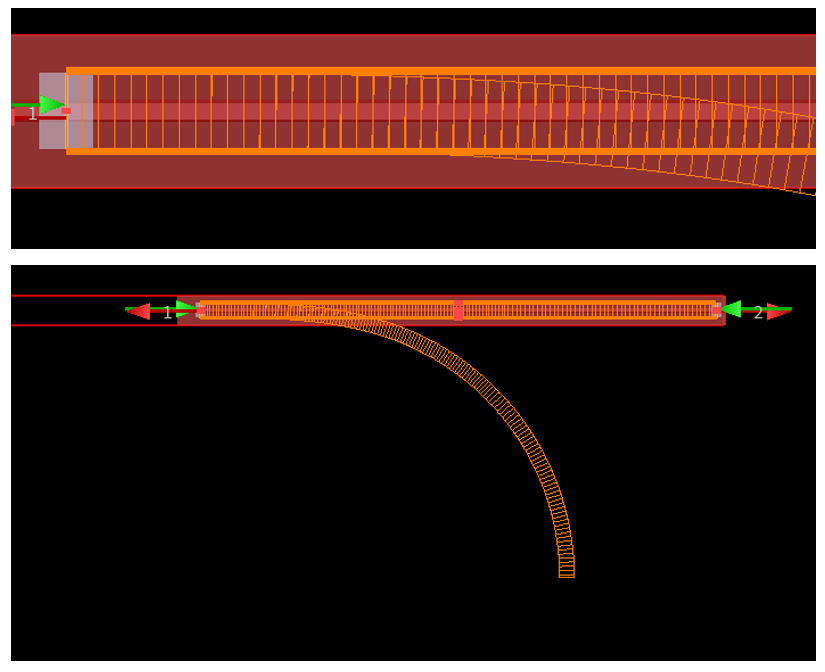}
    \caption{Discretized bend EME simulation geometry (screenshots taken in Lumerical MODE) used to run an approximate simulation to calculate waveguide bend scattering parameters at the pump wavelength. The top image shows the straight waveguide and the eigenmode expansion cells. The refractive index and mode monitor bend radius is set independently in each cell. The bottom image shows the simulation geometry overlaid with the ``real'' path the waveguide takes (curved waveguide).}\label{fig:bend-eme-sim}
\end{figure}

Designing the waveguide bends for our example resonator introduced a challenge of simulation scaling and accuracy. A high quality factor resonator ($10^5$--$10^6$) requires the waveguide bends to have high radii on platforms where the etch depth is comparable to the leftover substrate thickness, as propagation loss across bends increases as the bend radius decreases. On a 300-nm etch depth, 600-nm waveguide layer thickness (resulting in 300-nm-thick leftover substrate thickness) LiNbO\textsubscript{3} platform, bend radii of over \SI{100}{\micro\meter} are required to achieve sufficiently small losses. Furthermore, we are using a half-Euler-half-circular bend to reduce mode mismatch losses between the straight and bent waveguide sections. These result in a bend footprint of an approximately $150 \times 100$~\unit{\square\micro\meter} footprint for the half-bend (smallest unique cell, as the bend has 180$^\circ$ reflection symmetry). Two more complicating factors were that bends on an x-cut chip introduce anisotropy for TE-polarized modes and that dispersive material models introduce instabilities in FDTD simulations when the spatial grid is not dense enough. After compensating for these effects, the bend FDTD simulations at visible wavelengths (approximately \SI{780}{\nano\meter}) were too big to simulate on our machines within reasonable times that would permit iterating our design in a timely manner.

\begin{figure}[!t]
    \centering
    \includegraphics[width=1\linewidth]{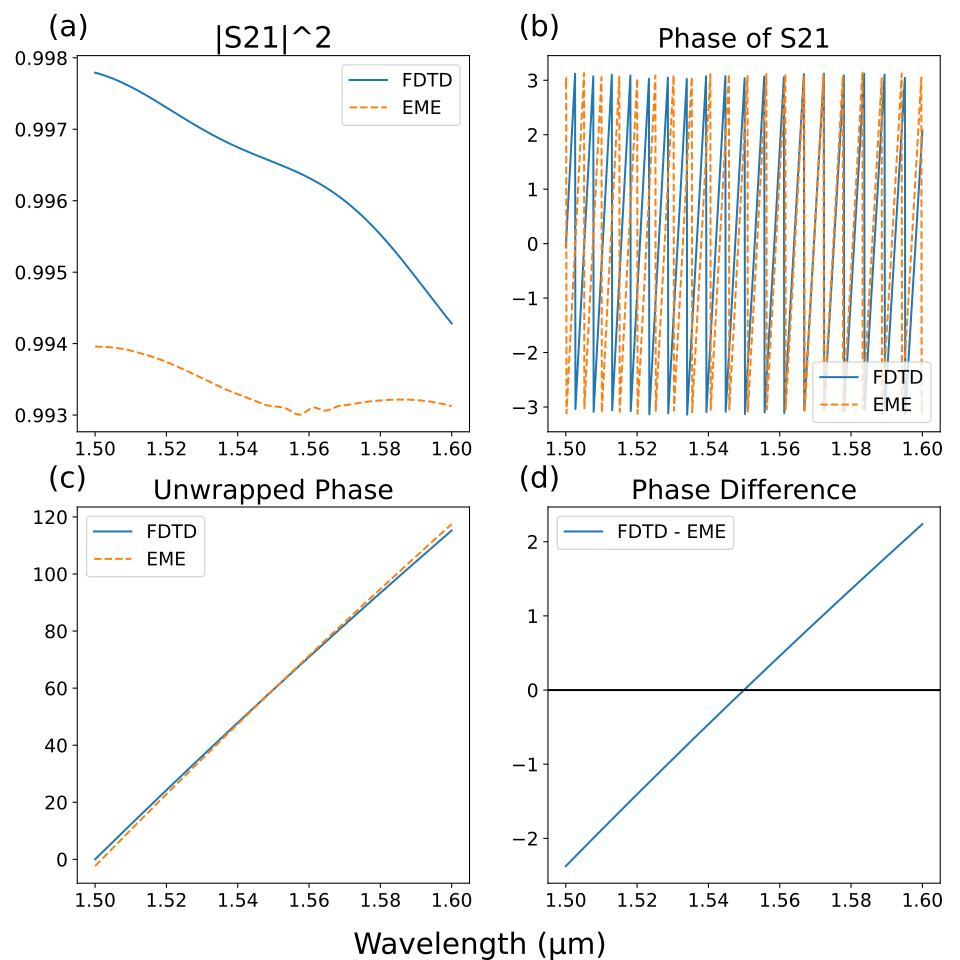}
    \caption{Comparison of waveguide bend scattering parameter amplitudes and phases. Plots (a) and (b) illustrate the amplitude squared and the phase of the bent waveguide transmission scattering parameter respectively. The phase accumulation of the waveguide vs wavelength is easier to analyze when the phase is unwrapped as in plot (c), and when the relative phase between the FDTD and sectioned EME methods is illustrated as in plot (d).}\label{fig:segmented-eme-vs-fdtd}
\end{figure}

We developed a segmented EME simulation technique that lets us approximate the scattering parameters of the waveguide bends by discretizing them into straight waveguide sections. The Lumerical MODE geometry is shown in Fig.~\ref{fig:bend-eme-sim} for this simulation configuration. We parametrize the waveguide curve and calculate the bend radius, section length, and tangential angle of each section. The bend radius and section lengths are used to define straight waveguide geometries for the EME simulation, while the tangential angle is used to interpolate the refractive index along the TE-polarized direction according to
\begin{equation}
    n(\phi,\omega) = \frac{n_1(\omega) n_2(\omega)}{\sqrt{n_1^2(\omega) \text{cos}^2(\phi) + n_2^2(\omega) \text{sin}^2(\phi)}}\text{,}
\end{equation}
where $\phi$ is the angle with respect to the in-plane ordinary axis. After performing the interpolation, we set the dispersive refractive index of the material for each waveguide segment to mimic an effective bent waveguide. We note that for x-cut devices, $n_1 = n_o\text{,}\ n_2 = n_e$, but since we utilize the TM modes in z-cut devices, the polarization is always out-of-plane of the chip and this interpolation is unnecessary.

We were able to compare FDTD and our segmented EME approximation for our waveguide bends at telecom wavelengths which were small enough to run FDTD simulations of. Figure~\ref{fig:segmented-eme-vs-fdtd} shows that the amplitude of the scattering parameters varies by less than 1\%. We note that the phase accumulation across the whole simulation bandwidth is inaccurate compared with FDTD, since a change of approximately 2.5 radians is a significant error. This error will affect the locations of resonances at our pump wavelengths, which means we cannot predict our pump detunings from the nearest resonance. However, the quality factors for each resonance should be mostly accurate---perhaps even an underestimate---compared with FDTD.

\bibliography{references}

\end{document}